\documentclass[a4paper,fleqn]{cas-sc}

\usepackage[numbers]{natbib}

\def\tsc#1{\csdef{#1}{\textsc{\lowercase{#1}}\xspace}}
\tsc{WGM}
\tsc{QE}
\tsc{EP}
\tsc{PMS}
\tsc{BEC}
\tsc{DE}
\begin{document}
\let\WriteBookmarks\relax
\def\floatpagepagefraction{1}
\def\textpagefraction{.001}
\shorttitle{L\'{e}vy noise-induced self-induced stochastic resonance in a memristive neuron}
\shortauthors{M.E Yamakou and T.D. Tran}

\title [mode = title]{L\'{e}vy noise-induced self-induced stochastic resonance  in a memristive neuron}

\author[1,3]{Marius E. Yamakou}[%[type=editor,
                        %auid=000,bioid=1,
                       % prefix=Sir,
                        %role=Researcher,
                        orcid=0000-0002-2809-1739
                        ]
\cormark[1]
\fnmark[1]
\ead{marius.yamakou@fau.de}
\address[1]{Department Mathematik, Friedrich-Alexander-Universit\"{a}t Erlangen-N\"{u}rnberg, Cauerstr. 11, 91058 Erlangen, Germany}

\author[2,3]{Tat Dat Tran}[%
orcid=0000-0001-5299-7023
]
\ead{tran@math.uni-leipzig.de}

\address[2]{Fakult\"{a}t f\"{u}r Mathematik und Informatik, Universit\"{a}t Leipzig, Augustusplatz 10, 04109 Leipzig, Germany}

\address[3]{Max-Planck-Institut f\"ur Mathematik in den Naturwissenschaften, Inselstr. 22, 04103 Leipzig, Germany}

\cortext[cor1]{Corresponding author}
\fntext[fn1]{This work was partially funded by the Deutsche Forschungsgemeinschaft (DFG, German Research Foundation): grant YA 764/1-1 to M.E.Y -- project number 456989199 and the Lehrstuhl f\"{u}r Angewandte Analysis (Alexander von Humboldt-Professur), Department Mathematik, Friedrich-Alexander-Universit\"{a}t Erlangen-N\"{u}rnberg, Germany.}

\begin{abstract}
Self-induced stochastic resonance (SISR) is a subtle resonance mechanism requiring a nontrivial scaling limit between the stochastic 
and the deterministic timescales of an excitable system, leading to the emergence of a limit cycle behavior which is absent without noise. 
All previous studies on SISR in neural systems have only considered the idealized Gaussian white noise.  Moreover, these studies have 
ignored one electrophysiological aspect of the nerve cell: its memristive properties. 
In this paper, first, we show that in the excitable regime, the asymptotic matching of the mean escape timescale of an $\alpha$-stable L\'{e}vy process (with value increasing as a power $\sigma^{-\alpha}$ of the noise amplitude $\sigma$, unlike the mean escape timescale of a Gaussian process with the value increasing as in Kramers' law) and the deterministic timescale (controlled by the singular parameter) can also induce a strong SISR.  In addition, it is shown that the degree of SISR induced by L\'{e}vy noise is not always higher than that of Gaussian noise. Second, we show that, for both types of noises, 
the two memristive properties of the neuron have opposite effects on the degree of SISR: 
the stronger the feedback gain parameter that controls the modulation of the membrane potential with the magnetic flux and the weaker the 
feedback gain parameter that controls the saturation of the magnetic flux, the higher the degree of SISR. 
Finally, we show that, for both types of noises, the degree of SISR in the memristive neuron is always higher than in the non-memristive neuron. 
Our results could find applications in designing  neuromorphic circuits operating in noisy regimes.
\end{abstract}


\maketitle

\section{Introduction}\label{Sec. I}
Noise is ubiquitous in neural systems and several studies have shown that it can play a constructive role in  
information processing 
\cite{wiesenfeld1995stochastic,douglass1993noise,guo2018functional,longtin1993stochastic,gammaitoni1998stochastic,wang2016levy,li2015stochastic,collins1996noise,nozaki1999effects}.
Noise-induced resonance mechanisms are a category of phenomena showing this constructive counter-intuitive role of noise. Several types of noise-induced 
resonance mechanisms have been identified and extensively studied, particularly in neural systems. These include stochastic resonance 
(SR) \cite{wiesenfeld1995stochastic,longtin1993stochastic,lindner2004effects,guo2017frequency,patel2008stochastic}, coherence resonance 
(CR) \cite{pikovsky1997coherence,gammaitoni1998stochastic,zhou2001array,neiman1997coherence,zhu2020phase}, spatial CR \cite{carrillo2004spatial,perc2005spatial},
inverse stochastic resonance \cite{gutkin2007transient,gutkin2009inhibition,yamakou2018weak,uzuntarla2013dynamical,yamakou2017simple}, 
recurrence resonance \cite{krauss2019recurrence}, and self-induced stochastic resonance 
(SISR) \cite{yamakou2017simple,muratov2005self,deville2005two,muratov2008noise,deville2007nontrivial,deville2007self,shen2010self,yamakou2018coherent,yamakou2019control,10.3389/fncom.2020.00062}. In this paper, we focus  on  SISR in a memristive neuron perturbed by a L\'{e}vy process -- a setting that has not been considered before.

SISR requires a nontrivial scaling limit between the stochastic and the deterministic timescales of an excitable system, leading to the emergence of quasi-periodic oscillations 
which are absent without noise. Generically, SISR occurs when a multiple-timescale excitable dynamical system is driven by a \textit{weak} noise amplitude. During SISR, the escape 
timescale of trajectories from one attracting region in phase space to another is distributed exponentially, and the associated transition rate is governed by an activation energy. 
Suppose the excitable system (e.g., a neuron) is placed out-of-equilibrium, and its activation energy decreases monotonically as the neuron relaxes slowly to a stable  quiescent state (stable fixed point); then, at a 
specific instant during the relaxation, the timescale of escape due to noise and the timescale of relaxation match, and the neuron fires at this point almost surely. 
If this activation brings the neuron back out-of-equilibrium, the relaxation stage can start over again, and the scenario repeats itself indefinitely, 
leading to a coherent spiking activity which cannot occur without noise. SISR essentially depends on the interplay of three different timescales: the slow and fast  timescales in the deterministic equation of the system, plus  a third timescale characteristic to the noise. 

It is important to note that the mechanism of SISR is very different from those of SR and CR. In fact, it has been shown in \cite{deville2005two} that CR and SISR are 
actually two distinct mechanisms even though both lead to the emergence of weak noise-induced coherent oscillations. Moreover, in our previous work \cite{yamakou2019control} 
(see also \cite{semenova2018weak}), it has been shown that the way SISR in the first layer of a duplex neural network controls CR in the second layer, is different from the 
control of CR when we have CR in the first layer. This difference in the controllability of CR by SISR and CR in multiplex networks  further confirms the fact that CR and 
SISR are actually different mechanisms. Compared to CR and SR, the conditions to be met for the mechanism of SISR are more subtle: Like CR, SISR does not require an external 
periodic signal as in SR. Remarkably, unlike CR, SISR does not require the system's parameters to be in the vicinity of  bifurcation thresholds, making it more robust to 
parametric perturbations than CR. Moreover, unlike both SR and CR, SISR requires a strong timescale separation between the variables of the excitable system.

All previous investigations on SISR have treated the input noise process as solely Gaussian \cite{yamakou2017simple,muratov2005self,deville2005two,muratov2008noise,deville2007nontrivial,deville2007self,shen2010self,yamakou2018coherent,yamakou2019control,10.3389/fncom.2020.00062}. But stochastic processes with a L\'{e}vy distribution are well-known to more accurately model the dynamics of real biological neurons \cite{wu2017levy,nurzaman2011levy}.
In general, dynamical systems composed of a large number of nonlinearly coupled subsystems often obey the L\'{e}vy distribution \cite{peng1993long,mantegna1995scaling,shlesinger1993strange}. Thus, in neural systems, the L\'{e}vy distribution on the network level reflects the emergent properties of the network in which the neurons are the subsystems. And at the level of the individual neuron, this implies that it is also composed of nonlinearly coupled subsystems -- the ionic channels.  In \cite{segev2002long}, a plot of interspike intervals and interevent intervals distributions indicates that neurons and neural network activities are characterized by a non-Gaussian heavy-tail interval distribution, thereby providing a solid reason as to why it makes sense to consider L\'{e}vy noise in the study of neural systems. L\'{e}vy noise has also been extensively used to model many other complex systems, including lasers \cite{rocha2020levy}, quantum dots \cite{novikov2005levy}, cardiac dynamics \cite{peng1993long}, molecular motor \cite{lisowski2015stepping}, economics \cite{stanley2000introduction,barndorff2001non}, and social systems \cite{perc2007transition}, where changes are often abrupt \cite{dubkov2008levy,xu2016switch}.

{Several studies on stochastic systems have departed from Gaussian to L\'{e}vy processes and compared their effects. For example, in \cite{perc2007transition}, the study of the stochastic  payoff  variations in  the spatial  prisoner’s  dilemma  game is presented; in \cite{feng2019effects}, the neuron competition models; and in \cite{guo2021dynamical}, the statistical complexity and normalized Shannon entropy of the FitzHugh–Nagumo neuron model. In this paper, in a similar fashion, we study SISR in a memristive neuron perturbed by a  L\'{e}vy white noise.  The analytical conditions required for the occurrence of SISR and the parameters combination of the L\'{e}vy noise that maximize the degree of SISR are obtained. Then, we compare these analytical conditions and the degree of SISR when it is induced by Gaussian noise.}

The exchange of charged ions across the membrane of the nerve cell can induce complex electromagnetic field inside and outside this membrane, and the 
membrane potential of neuron gets modulated by the induced electromagnetic field. Thus, by Faraday's law of electromagnetic induction, the effect of electromagnetic 
induction on the cell must be considered. Recently, M. Lv et al. \cite{lv2016model} proposed a modified neural model that takes into account the effect of 
the magnetic field generated by the internal bioelectricity of the nerve cell (i.e., the movement of charged ions  across the membrane on the spiking activity of the cell).
 In the modified (improved) neuron models, the effects of electromagnetic induction are described by using the magnetic flux. And the modulation of the 
 membrane potential by the magnetic flux is realized by using a memristor coupling, hence the term memristive neurons \cite{chua1971memristor}. The modification of the 
 original neural models, so that they take into account these electromagnetic effects, consisted of adding a variable for the magnetic flux into the original equations.
 
Several studies have shown that memristive neurons can generate a rich variety of modes in electric activities by not only varying the external input current, 
but also by varying the magnetic flux parameters --- those that control the memristive properties of the neuron 
\cite{Yamakou2020,wu2017dynamical,ma2017phase,xu2018collective,ma2017electromagnetic,wu2016model}. 
It has been shown that the magnetic flux coupling between neurons can induce perfect phase synchronization of chaotic time series of membrane potentials \cite{ma2017phase}. 
This result basically showed that neurons exposed to their own external magnetic field can induce phase synchronization and appropriate behaviors can be selected from different 
magnetic flux parameter values. 

It has also been shown that the magnetic field coupling can contribute to the signal exchange between neurons by triggering superposition of electric field when synapse 
coupling is not available \cite{xu2018collective}. Here, the contribution of field coupling from each neuron is described by introducing appropriate weight dependent on 
the distance between two neurons. It was found that the degree of synchronization is dependent on the intensity and weight of the field coupling and that the pattern 
selection of the network connected with gap junction can be modulated by this field coupling. 

The memristive properties have also been shown to play a significant role in the dynamics of other types of biological tissues. For example, it  has been shown  that 
target  wave  propagation can be blocked to stand in a local area of the cardiac tissue and the excitability of this tissue can be suppressed to approach quiescent 
but homogeneous state when electromagnetic flux (generated by the motion of ions across the membrane of the cardiac cell) is imposed on the cardiac tissue \cite{ma2017electromagnetic}. 
Moreover, it has been shown that a spiral wave can be triggered and developed by setting specific initial conditions in the cardiac tissue under the effects 
of magnetic flux, i.e., the tissue still support the survival of standing spiral waves under specific values of the magnetic flux parameters \cite{wu2016model}.

It is now well-accepted that the effects of the magnetic flux across the membrane of the cell should be considered when investigating the emergence of electrical activities and wave propagation in the nerve and cardiac cells \cite{lv2016model,ma2017electromagnetic}. 
However, all previous studies on SISR in neural systems have been done only with non-memristive models perturbed by Gaussian noise. Thus, the effect of the memristive properties of a neuron on L\'{e}vy and Gaussian noise-induced SISR are still unknown. In this paper, we bridge this gap by applying nonlinear dynamics methods and numerical simulations to address the following questions: (i) Can L\'{e}vy noise (with polynomial intrinsic timescale) also induce SISR? (ii) Which noise induces the 
highest degree of SISR, L\'{e}vy or Gaussian noise? (iii) How do the memristive properties of the neuron affect the degree of SISR induced by these two types of noises?

The rest of the paper is organized as follows: In section~\eqref{Sec. II}, we describe the mathematical equation modelling a memristive neuron driven by L\'{e}vy noise and 
we also determine the excitable parameter space of model in terms of the memristive parameters.
Section~\eqref{Sec. III} is devoted to the theoretical analysis of the mechanism of SISR.
In section~\eqref{Sec. IV}, we present and discuss the numerical results. And in Section~\eqref{Sec. V}, we have summary and conclusions.

\section{Mathematical model and excitability}\label{Sec. II}
\subsection{Model description}
We consider a memristive FitzHugh-Nagumo (FHN) neuron model of type-II excitability \cite{lv2016model,fitzhugh1969mathematical}, 
driven by an $\alpha$-stable L\'{e}vy process, and described by the following stochastic differential equations
\begin{equation}\label{eq:1}
\begin{split}
\left\{\begin{array}{lcl}
dv_{\tau}  &=&  \varepsilon^{-1}f_1(v_{\tau},w_{\tau},\phi_{\tau})d\tau + \frac{1}{\sqrt[\alpha]{\varepsilon}} dL^{\alpha,\beta}(\tau;\sigma,\mu),\\[3.0mm]
dw_{\tau}  &=&  f_2(v_{\tau},w_{\tau},\phi_{\tau})d\tau,\\[3.0mm]
d\phi_{\tau}  &=&  f_3(v_{\tau},w_{\tau},\phi_{\tau})d\tau,
\end{array}\right.
\end{split}
\end{equation}
with the deterministic velocity vector field given by
\begin{equation}\label{eq:2}
\begin{split}
\left\{\begin{array}{lcl}
f_1(v,w,\phi)&=&\displaystyle v-\frac{v^3}{3}-w - k_1\rho(\phi)v,\\[3.0mm]
f_2(v,w,\phi) &=& v + d - c w,\\[3.0mm]
f_3(v,w,\phi) &=& v - k_2\phi,
\end{array}\right.
\end{split}
\end{equation}
where $(v,w,\phi)\in\mathbb{R}^3$ represent the action potential variable $v$, the recovery current (or sodium gating) variable $w$ 
that restores the resting state of the neuron, and the third variable $\phi$ is the magnetic flux across membrane which can generate additive current.

The parameter $0<\varepsilon:=\tau/t\ll1$ is timescale separation ratio (also called singular parameter) between the slow timescale $\tau$ and the fast timescale $t$. It 
accounts for the slow kinetics of the sodium channel in the nerve cell and controls the main morphology of the action potential generated \cite{xu2014parameters}. 
It is worth noting that $\varepsilon$ is a very small and positive parameter ($0<\varepsilon\ll1$), and from Geometric Singular Perturbation Theory (GSPT) for slow-fast dynamical systems in the standard form \cite{kuehn2015spatial}, this means that the $v$-variable is fast and the $w$- and $\phi$-variables are slow. Moreover, from GSPT, the relation $\varepsilon:=\tau/t$ can be used (i.e., $d\tau:=\varepsilon dt$) to transform the Eq.~\eqref{eq:1} from the slow timescale $\tau$ to the fast timescale $t$, given by Eq.~\eqref{eq:stoch}. We further note that Eq.~\eqref{eq:1} and Eq.~\eqref{eq:stoch} are equivalent except that their orbits evolve on different timescales.The constant parameter $d$ is such $d\in(0,1)$, and $c>0$ is a codimension-one Hopf bifurcation parameter. 

The term $\rho(\phi)$ in Eq.~\eqref{eq:2} is the memory conductance of a magnetic flux-controlled memristor and it is used to describe the coupling
between magnetic flux $\phi$ and membrane potential $v$ of the neuron \cite{wu2019new,bao2010steady,muthuswamy2010implementing}.
The memory conductance of a memristor is often described by
\begin{equation}\label{eq:3}
\rho(\phi) = a + 3b \phi^2,
\end{equation}
where $a$ and $b$ are constant parameters. In this paper, we fix $a=0.1$ and $b=0.02$, to stay consistent with other works \cite{li2015hyperchaos}.
The magnetic feedback gain parameters $k_1$ and $k_2$ describe the interaction between the magnetic flux and membrane potential.
More precisely, $k_1$ bridges the coupling and modulation on the membrane potential $v$ from magnetic flux $\phi$, 
and $k_2$ describes the degree of polarization and magnetization by adjusting the saturation of magnetic flux \cite{ma2017mode}.
The term $k_1\rho(\phi)v$ in Eq.~\eqref{eq:2}, therefore, describes the  modulation on the membrane potential of the neuron, 
and it depends on the variation in the magnetic flux. 
Combining Faraday's law of electromagnetic induction and the basic properties of a memristor, the term  $k_1\rho(\phi)v$ is regarded as additive induction current on the membrane potential.
The dependence of electric charge $q$ on the magnetic flux $\phi$ is defined as \cite{hong2013design}
\begin{equation}\label{eq:4}
\rho(\phi)=\frac{dq(\phi)}{d\phi} = a + 3b \phi^2.
\end{equation}
Moreover, because the current $i$ is defined as the time derivative of charge $q$, the physical significance for the term $\rho(\phi)v$ could be described as
\begin{equation}\label{eq:5} 
i=\frac{dq(\phi)}{dt}=\frac{dq(\phi)}{d\phi}\frac{d\phi}{dt}=\rho(\phi)V=k_1\rho(\phi)v,
\end{equation}
where $V$ denotes an induced electromotive force with a feedback gain parameter $k_1$.
The potassium and sodium ionic currents contribute to the magnetic flux across the membrane and also to the membrane potential. This introduces a negative feedback term $-k_2\phi$ in 
the third equation of Eq.~\eqref{eq:2}.

\begin{figure}%[htp!]
\begin{center}
\includegraphics[width=8.0cm,height=5.0cm]{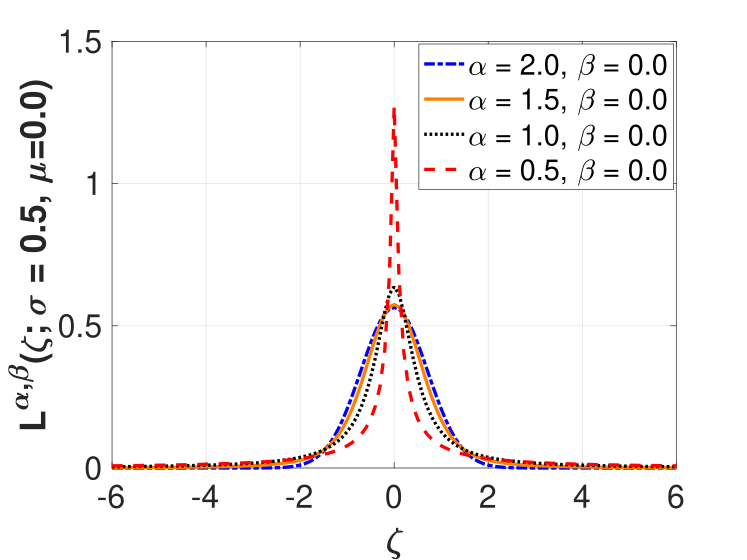} \includegraphics[width=8.0cm,height=5.0cm]{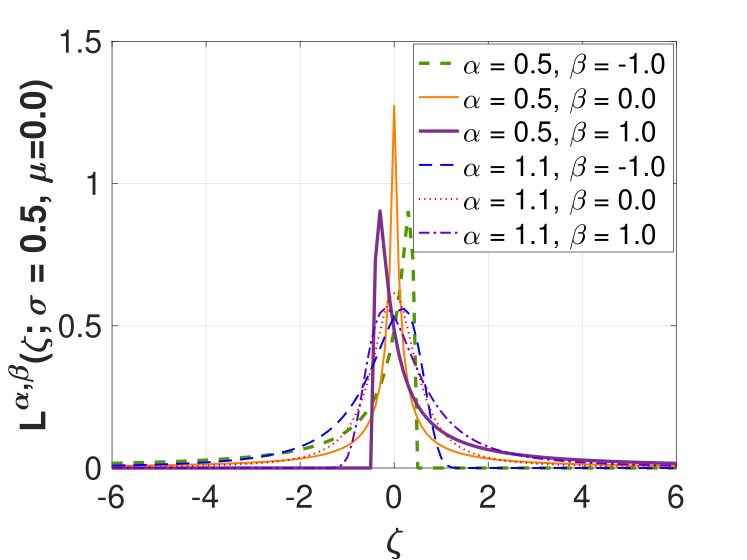}
\end{center}
\caption{Probability density functions for L\'evy distribution 
of $L^{\alpha,\beta}(\zeta; \sigma=0.5, \mu=0.0)$ with different values of the stability index and skewness parameters.}\label{fig:1}
\end{figure}

$L^{\alpha,\beta}(\tau;\sigma,\mu)$ is an independent $\alpha$-stable L\'{e}vy motion. The L\'{e}vy motion, as an appropriate model for non-Gaussian
processes with jumps \cite{sato1999levy,bertoin1996levy}, has properties of stationary and independent increments.
Throughout this paper, we adhere to one of possible parametrizations of $\alpha$-stable distributions \cite{dybiec2009levy,dybiec2007escape,dybiec2009levy,prokhorov1965w} 
which allows to write down the characteristic function of an appropriate probability distribution
\begin{equation}\label{eq:6}
 \phi(x)=\int_{-\infty}^{\infty} e^{-ix\zeta} L^{\alpha,\beta}(\zeta;\sigma,\mu)d\zeta,
\end{equation}
in the form of 
\begin{equation}\label{eq:7}
\phi(x)=\exp\Big[i\mu x -\sigma^{\alpha}|x|^{\alpha}\Big(1-i\beta sgn(x)\tan\frac{\pi\alpha}{2}\Big)\Big],
\end{equation}
if $\alpha\in(0,1)\bigcup(1,2]$, or 
\begin{equation}\label{eq:8}
\phi(x)=\exp\Big[i\mu x -\sigma|x|\Big(1 + i\beta\frac{2}{\pi}sgn(x)\ln|x|\Big)\Big],
\end{equation}
if $\alpha=1$. Here, $\alpha$ stands for the stability index and lies in the interval $\alpha\in(0, 2]$. It describes an asymptotic power law of the $\zeta$-distribution, 
$L^{\alpha,\beta}(\zeta;\sigma,\mu )\sim|\zeta|^{-(\alpha+1)}$, and controls the impulsiveness (i.e., the jump frequency and size) of the process. The parameter $\beta\in[-1,1]$ determines the skewness (asymmetry) of the distribution.
{$\sigma\in(0,\infty)$ is the scale parameter. $\mu\in\mathbb{R}$ is the location parameter.} {Closed, analytical forms of the stable L\'evy probability densities are known in some cases. For example, $L^{2,0}(\cdot;\sigma,\mu)$ is the well-known Gaussian distribution; $L^{1,0}(\cdot;\sigma,\mu)$ yields the Cauchy distribution; $L^{\frac{1}{2},1}(\cdot;\sigma,\mu)$ yields the L\'{e}vy-Smirnoff ($\zeta > \mu$) distribution; and other forms can be found in \cite{penson2010exact,gorska2011levy}. 

Fig.~\ref{fig:1} shows the probability density functions of Lévy distribution
$L^{\alpha,\beta}(\zeta; \sigma, \mu)$ with some values of the stability index and 
skewness parameters. Throughout this paper, we fix the location parameter at $\mu = 0.0$ and use interchangeably notations $L^{\alpha,\beta}(\zeta), L(\zeta)$, and $L_{\zeta}$.}

\subsection{The excitable regime of the model}
The deterministic memristive FHN neuron (i.e., Eq.~\eqref{eq:1} without the noise term) with a \textit{unique} and \textit{stable} fixed point cannot maintain a self-sustained spiking activity. 
One says in this case that the neuron is in the excitable regime \cite{izhikevich2000neural}, in contrast to the oscillatory regime, where the neuron continuously spikes due to 
the occurrence of a bifurcation onto a limit cycle. 
In the excitable regime, choosing an initial condition in the basin of attraction of this unique and stable fixed point will result in at \textit{most one} large non-monotonic 
excursion into the phase space after which the trajectory returns to this fixed point and stays there until the initial conditions are changed again. 

The deterministic predisposition required for SISR is an excitable regime, so that during SISR, the self-sustained and coherent spike trains 
produced by the neuron is due \textit{only} to the presence of noise and not because of the occurrence of bifurcations onto a limit cycle. This is one of the crucial 
differences between SISR and CR --- the predisposition required for the latter mechanism is the close proximity of parameters to the bifurcation threshold, so that weak noise 
amplitudes can easily drive the system to this bifurcation threshold without, stochastically, overwhelming the dynamics \cite{pikovsky1997coherence,neiman1997coherence,deville2005two}.

In this subsection, we determine the excitable regime of the memristive FHN neuron model in terms of the Hopf bifurcation and memristive parameters. At the fixed points $(v_e,w_e,\phi_e) \in Fix$ (the set of rest states of the neuron), the variables
$v(\tau)$, $w(\tau)$, and $\phi(\tau)$ reach a stationary state, while the set of
fixed points defined by the intersection of the nullclines as
\begin{equation}\label{eq:9}
Fix :=\left \{ (v,w,\phi) \in \mathbb{R}^3: f_1=f_2=f_3=0 \right\},
\end{equation}
depends on the parameters $c$, $d$, $k_1$, and $k_2$. The sign of
\begin{equation}\label{eq:10}
\Delta=\frac{g^2}{4} + \frac{p^3}{27} ,
\end{equation}
determines the number of fixed points.
In this paper, we consider the case where we have only one stable fixed point.
If $\Delta>0$, we have a unique fixed point given by
\begin{equation}\label{eq:11}
\begin{split}
\left\{\begin{array}{lcl}
\displaystyle{v_{e}=\sqrt[3]{-\frac{g}{2}-\sqrt{\Delta}}+\sqrt[3]{-\frac{g}{2}+\sqrt{\Delta}}} \\[4.0mm]
 \displaystyle{w_{e}=\frac{1}{c}(v_{e}+d)},\\[4.0mm]
 \displaystyle{\phi_e= \frac{v_e}{k_2}},
\end{array}\right.
\end{split}
\end{equation}
where
\begin{equation}\label{eq:12}
\begin{split}
\left\{\begin{array}{lcl}
p = \displaystyle{ \frac{\frac{1}{c} + k_1 a -1}{\frac{1}{3} + \frac{3 k_1 b}{k_2^2}}},\\[8.0mm]
g = \displaystyle{ \frac{\frac{d}{c}}{\frac{1}{3} + \frac{3 k_1 b}{k_2^2}}}.
\end{array}\right.
\end{split}
\end{equation}
 
Moreover, in the model we arbitrarily fix $d=0.5$ once and for all, and we determine the excitable regime of the model in terms of the parameter $c$ and the two new 
parameters $k_1$ and $k_2$ --- also known as the magnetic gain parameters.
  With the fixed values of the parameters $a=0.1$, $b=0.02$, and $d=0.5$, $p$ and $g$ in  Eq.~\eqref{eq:12} now depend only on $c$, $k_1$, and $k_2$. We have: 
\begin{equation}
\begin{split}
\left\{\begin{array}{lcl}
p = \displaystyle{\frac{-1 + \frac{1}{c} + 0.1 k_1}{\frac{1}{3} + \frac{0.06 k_1}{k_2^2}}},\\[7.0mm]
g = \displaystyle{\frac{0.5}{c\Big(\frac{1}{3} + \frac{0.06 k_1}{k_2^2}\Big)}},
\end{array}\right.
\end{split}
\end{equation}
 which are both always positive for $c<1, k_1\geq0$ and $k_2>0$. Hence, $\Delta$ in  Eq.~\eqref{eq:10} will always be positive for $c<1, k_1\geq0$ and $k_2>0$, ensuring 
the uniqueness of the fixed point $(v_{e}, w_{e},\phi_{e})$ in Eq.~\eqref{eq:11}.

\begin{figure}%[htp!]
\begin{center}
\includegraphics[width=8.0cm,height=5.0cm]{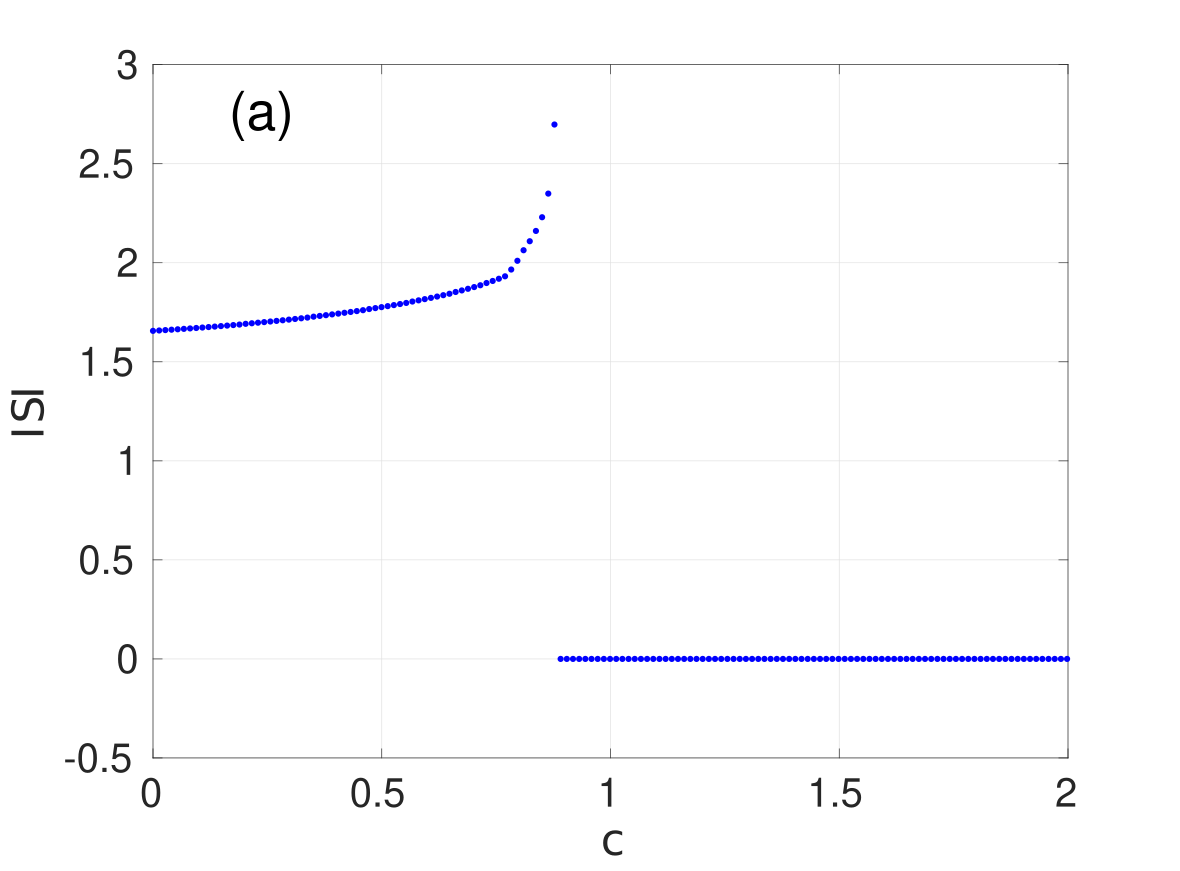} \includegraphics[width=8.0cm,height=5.0cm]{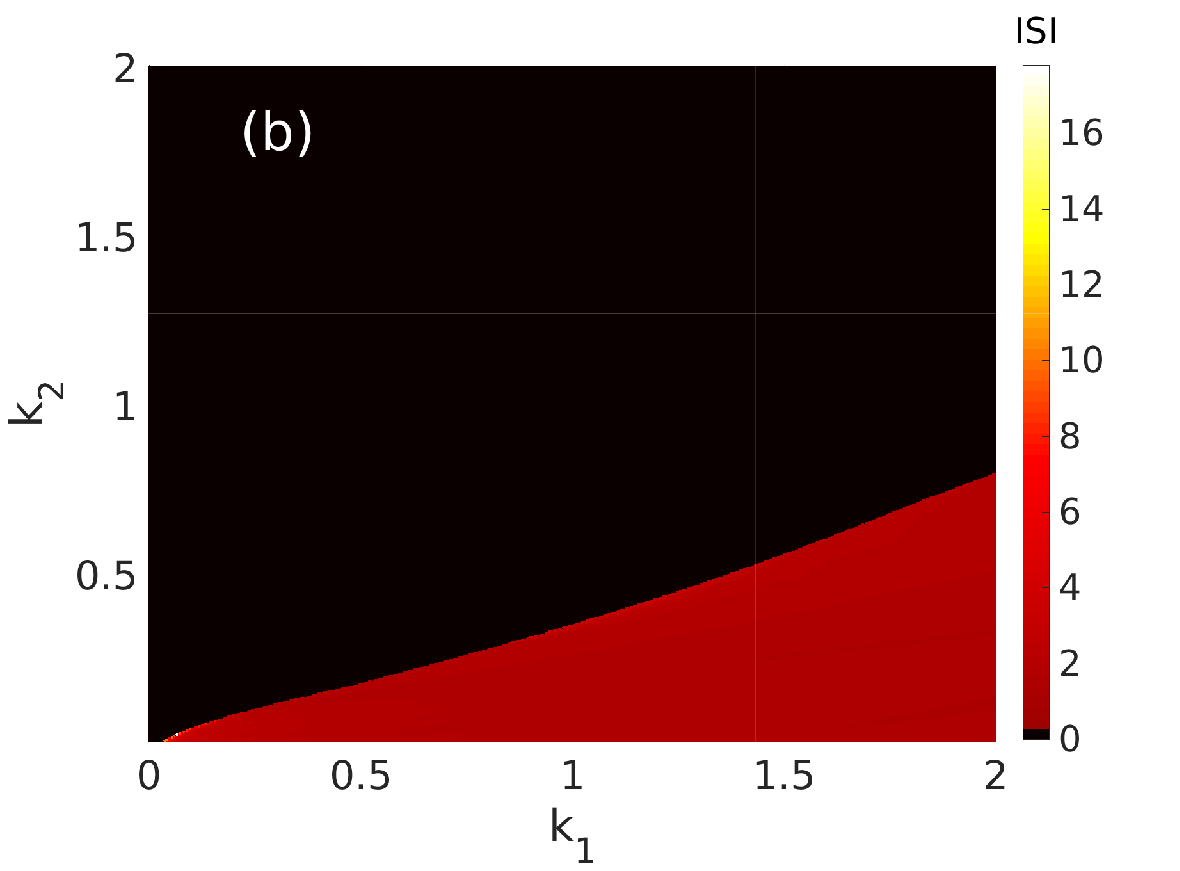}
\end{center}
\caption{Panel \textbf{(a)}: Bifurcation diagram with respect to parameter $c$, showing the oscillatory ($ISI>0$) and excitable ($ISI=0$) regimes in $c<c_h=0.875$ and $c\geq c_h$, respectively,
with $k_1=0.1$ and $k_2=0.1$. Panel \textbf{(b)}: Color-coded $ISI$ for a two-parameter space bifurcation diagram with respect to $k_1$ and $k_2$ at $c=0.95>c_h$, 
showing the oscillatory regime in red and yellow where $ISI>0$ and the excitable regime in dark where $ISI=0$. 
In both panels, the other parameter values are fixed at: $a=0.1$, $b=0.02$, $d=0.5$, and $\varepsilon=0.001$.}\label{fig:2}
\end{figure}

With initial conditions at the unique fixed point $\big[v_{e}(c,k_1,k_2), w_{e}(c,k_1,k_2),\phi_{e}(c,k_1,k_2)\big]$, we numerically computed 
a codimension-one and codimension-two bifurcations, showing the excitable and oscillatory regimes of the memristive neuron with the respect to the  parameter $c$ in 
Fig.~\ref{fig:2}\textbf{(a)} and the magnetic gain parameters $k_1$ and $k_2$ in Fig.~\ref{fig:2}\textbf{(b)}, respectively. 

The bifurcation diagram in Fig.~\ref{fig:2}\textbf{(a)} shows a non-zero inter-spike interval ($ISI$) for $0<c<c_h$, where $c_h=0.875$ is the super-critical Hopf bifurcation threshold.
For $c\geq c_h$, there is no spiking, i.e., $ISI=0$, indicating that the neuron is in an excitable regime at $k_1=0.1$ and $k_2=0.1$. 
However, it is well-known that 
variations in these magnetic gain parameters can significantly affect the dynamical response of the neuron \cite{ma2017mode}, thereby switching the neuron's dynamics from an excitable 
to an oscillatory regime and vice versa, even when $c>c_h$. Hence, it is important to determine the range of values of $k_1$ and $k_2$ in which the neuron will remain in the 
excitable regime for a particular value of $c$, chosen such that $c_h<c<1$.

Fig.~\ref{fig:2}\textbf{(b)} shows, for $c=0.95>c_h=0.875$ (i.e., $c$ is far enough from the bifurcation threshold and also less than one so that the stable fixed point is unique), 
a two-parameter space bifurcation diagram with respect to $k_1$ and $k_2$.  We also note that $k_2$ starts at a non-zero value, i.e., at $k_2=0.01$, to ensure that our fixed point 
in Eq.~\eqref{eq:11} is unique. 
The color-coded $ISI$ shows the oscillatory regime in red and yellow where $ISI>0$. The yellow region corresponds to few points around the origin of the $(k_1,k_2)$ plane, where
$ISI$ takes relatively large values. For example, at $k_1=0.0361$ and $k_2=0.01$ we have $ISI=10.16$, and at $k_1=0.0643$ and $k_2=0.03$, $ISI$ takes its largest value, i.e., $ISI=17.78$.
The  dark region (where $ISI=0$) corresponds to the excitable regime, with the deterministic model in Eq.~\eqref{eq:1} 
consisting of unique and stable fixed point given by Eq.~\eqref{eq:11}. 
Therefore, throughout this paper, we will investigate the mechanism of SISR when the neuron is in the excitable regime defined by: 
$c=0.95$, $k_1\in[0.0,2.0]$, $k_2\in[1.0,2.0]$, $a=0.1$, $b=0.02$, $d=0.5$, and $\varepsilon=0.001\ll1$.

\section{The asymptotic matching of timescales and SISR} \label{Sec. III}
Now we consider Eq.~\eqref{eq:1} such that its deterministic version is in the excitable regime, defined by the parameters intervals and values above. 
To understand how noise can induced a regular escape of trajectories from the basin of attraction of the stable fixed point, leading to the emergence of a 
coherent spike train, we transform Eq.~\eqref{eq:1} from the slow timescale $\tau$ to the fast timescale $t$ to obtain Eq.~\eqref{eq:stoch} {using the relation $\varepsilon:=\tau/t$ or more precisely, $d\tau=\varepsilon dt$ \cite{kuehn2015spatial}}. Under this timescale transformation the noise term is re-scaled according to the scaling law of L\'evy motion. 
That is, if $L_{\tau}$ is a L\'evy motion, then for every $\lambda>0, \lambda^{-\frac{1}{\alpha}} L_{\lambda \tau}$ is also a L\'{e}vy motion 
(i.e., they have the same distribution). {Furthermore, we consider the standard form of the L\'evy noise, i.e., $L^{\alpha,\beta}(\tau;\sigma,0)=\sigma \hat{L}^{\alpha,\beta}(\tau;1,0)$, where the scale parameter $\sigma$ clearly represents the noise intensity.} We note that because of this scaling law, 
the term $1/\sqrt[\alpha]{\varepsilon}$ was introduced in the noise term in Eq.~\eqref{eq:1} to guarantee that
in Eq.~\eqref{eq:stoch}, the noise intensity, $\sigma$, measures the relative strength of the noise term compared to the 
deterministic term $f_1(v_{t},w_{t},\phi_{t})$ irrespective of the value of $\varepsilon$.
\begin{equation}\label{eq:stoch}
\begin{split}
\left\{\begin{array}{lcl}
dv_{t} & = & f_1(v_{t},w_{t},\phi_{t})dt + \sigma d\hat{L}_{t}^{\alpha,\beta},\\[3.0mm]
dw_{t} & = &\varepsilon f_2(v_{t},w_{t},\phi_{t})dt,\\[3.0mm]
d\phi_{t} & = &\varepsilon f_3(v_{t},w_{t},\phi_{t})dt.
\end{array}\right.
\end{split}
\end{equation}

\begin{figure}%[htb!]
\begin{center}
\includegraphics[width=5.5cm,height=5.0cm]{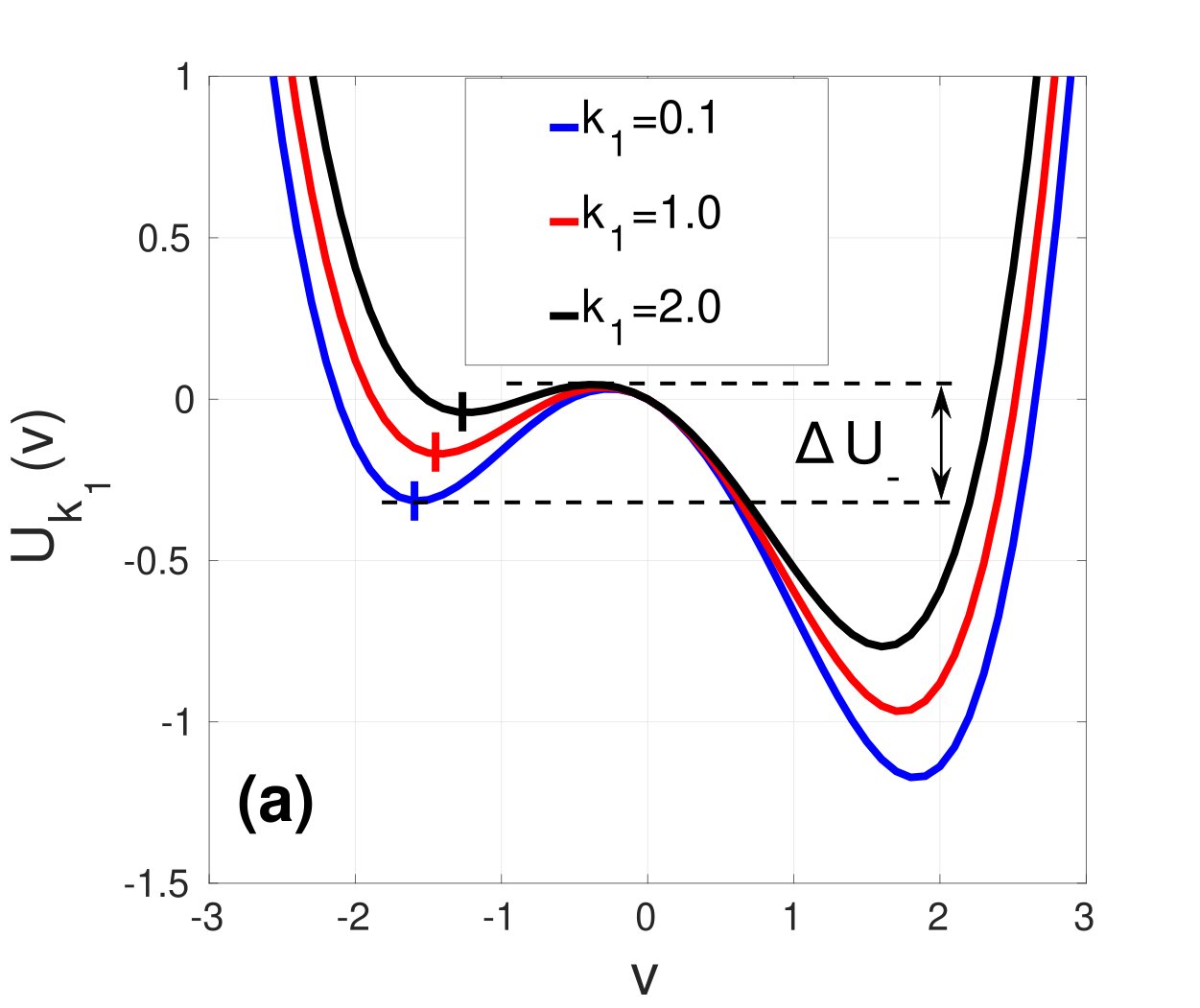}\includegraphics[width=5.5cm,height=5.0cm]{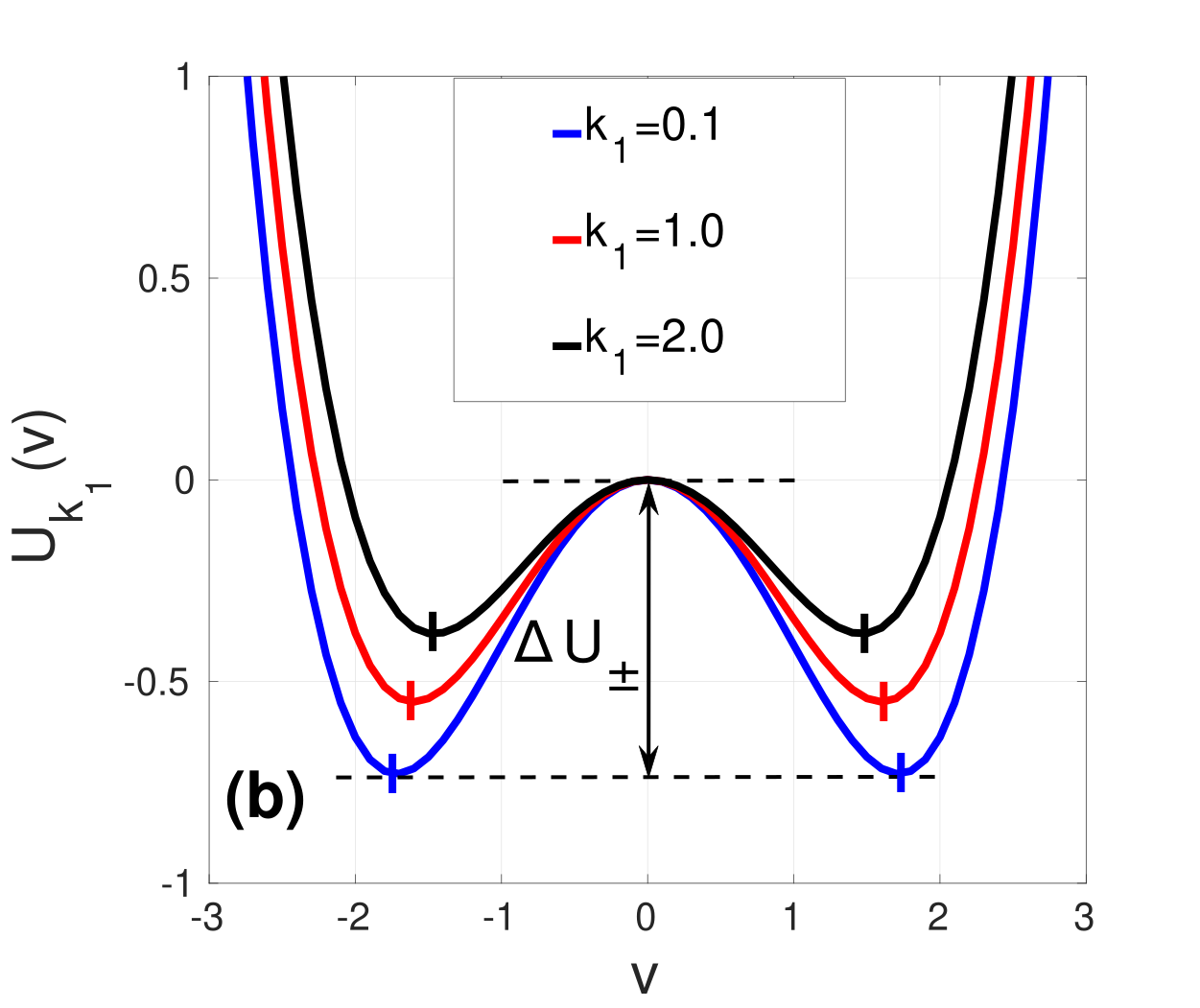}\includegraphics[width=5.5cm,height=5.0cm]{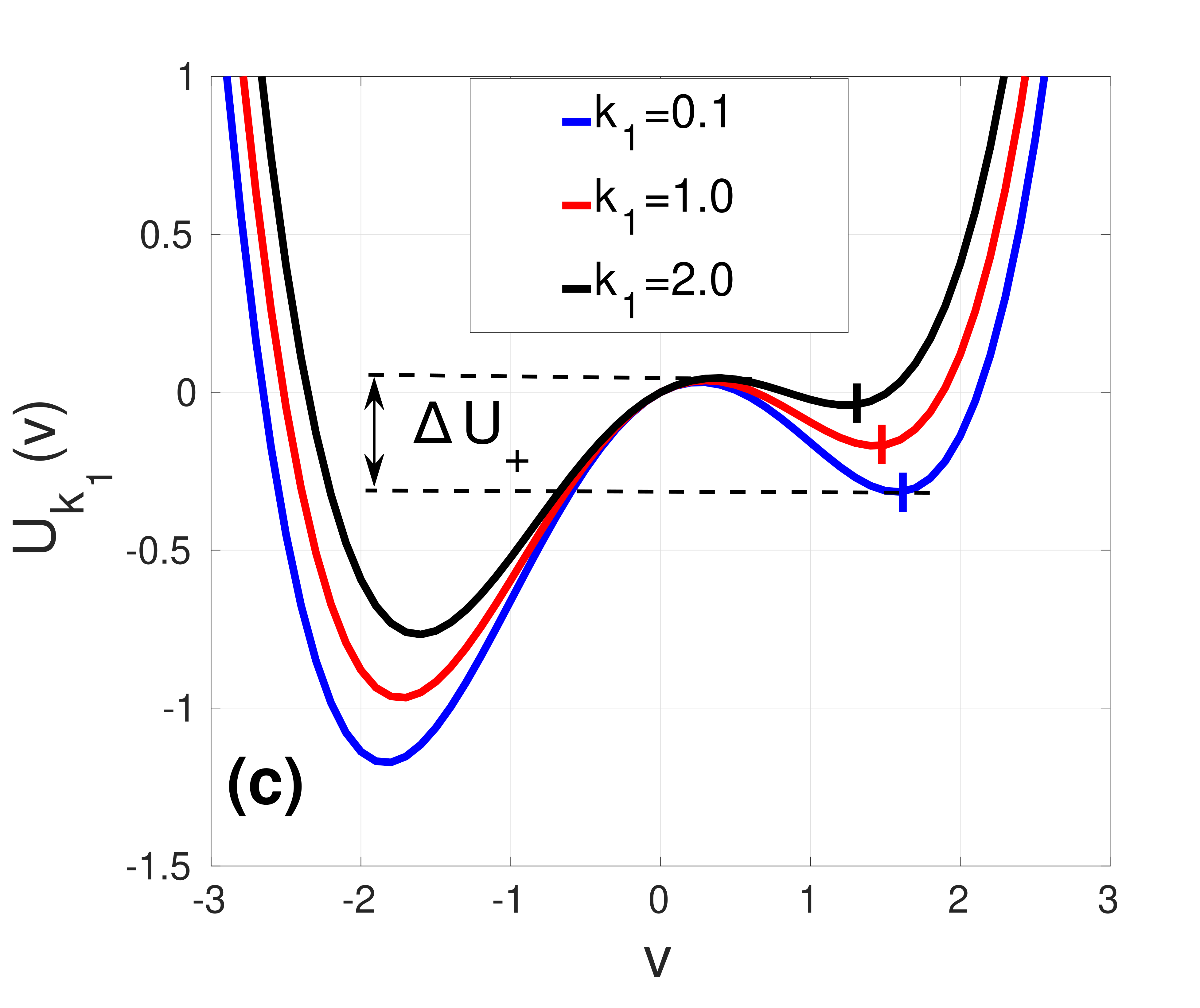}
\end{center}
\caption{{Variations of the potential $U_{k_1}(v)$ given in Eq.~\eqref{pot}. The energy barriers $\bigtriangleup U_{\pm}$ are indicated in the asymmetric 
cases in \textbf{(a)} ($w<0$) and in \textbf{(c)} ($w>0$), and in the symmetric case in \textbf{(b)} ($w=0$).
The band widths of the wells are given by the distances between the minima located at $v_l$ and $v_r$ (short vertical bars) and the saddle point located at $v=v_m=0$. The stronger the magnetic gain parameter $k_1$, the shallower 
the energy barriers $\bigtriangleup U_{\pm}$ and the shorter the band widths. In \textbf{(a)}, $w=-0.25$, in \textbf{(c)} $w=0.25$, and in all panels $\phi= 0.85$.}}\label{fig:3}
\end{figure}

In the adiabatic limit $\varepsilon \to 0$, the timescale separation between $v_t$ and the two other variables 
$w_t$ and $\phi_t$ become very large. This indicates that $w_t$ and $\phi_t$ are frozen on the $O(1)$ fast timescale. Hence, Eq.~\eqref{eq:stoch} is approximated by Eq.~\eqref{eq:stoch1} 
\begin{equation}\label{eq:stoch1} %Eq. 14
\begin{split}
\left\{\begin{array}{lcl}
dv_t &=& - U_{k_1}'(v_{t})dt +  \sigma d\hat{L}^{\alpha,\beta}_{t},\\
dw_t &=& 0,\\
d\phi_t &=& 0,\\
\end{array}\right.
\end{split}
\end{equation}
where $U_{k_1}'(v_{t})$ is the derivative of the potential 
\begin{equation}\label{pot} %Eq. 15
U_{k_1}(v) = \frac{1}{12}v^4  - \frac{1- k_1 \rho(\phi)}{2} v^2 + w v,
\end{equation} 
with respect $v$. $U_{k_1}(v)$ is the double-well potential with the constant solutions of the last two equations in Eq.~\eqref{eq:stoch1} given by $w$ and $\phi$, 
respectively. This potential has, respectively, a left local minimum, a saddle, and a right local minimum at
\begin{equation}\label{eq:sol} %Eq. 16
\begin{split}
\left\{\begin{array}{lcl}
v_{l} & = &2 \sqrt{-\frac{P}{3}} \cos\Bigg(\frac{1}{3} \arccos\Bigg(\frac{3Q}{2P} \sqrt{\frac{-3}{P}}\Bigg) + \frac{2\pi}{3}\Bigg),\\[4.0mm]
v_{m} & = &2 \sqrt{-\frac{P}{3}} \cos\Bigg(\frac{1}{3} \arccos\Bigg(\frac{3Q}{2P} \sqrt{\frac{-3}{P}}\Bigg) - \frac{2\pi}{3}\Bigg),\\[4.0mm]
v_{r} & = &2 \sqrt{-\frac{P}{3}} \cos\Bigg(\frac{1}{3} \arccos\Bigg(\frac{3Q}{2P} \sqrt{\frac{-3}{P}}\Bigg)\Bigg),
\end{array}\right.
\end{split}
\end{equation}
where $P = 3 [ k_1 \rho(\phi)- 1]$ and $Q = 3 w$, see Fig.~\ref{fig:3}.

It was shown in \cite{Chechkin07,Imkeller06} that for barrier crossing phenomena driven by L\'evy white noise in the double-well potential, the mean exit time 
from one of the wells increases as a power $\sigma^{-\alpha}$ of the noise intensity $\sigma$ with 
$\sigma \to 0$ and not exponentially as with Gaussian white noise would do in the limit 
as $\sigma \to 0$ \cite{yamakou2018coherent,kramers1940brownian,stambaugh2006noise}. By applying the general results presented in~\cite{Imkeller06} to our particular case,  we calculated for the double-well potential in Eq.~\eqref{pot}, the mean exit times of the L\'{e}vy process as: 
\begin{equation}\label{eq:sol1}%Eq, 17
\begin{split}
\left\{\begin{array}{lcl}
\displaystyle{\mathbb E T_{exit}(v_l \to v_r) \approx \frac{\alpha |v_{l}|^{\alpha}}{\sigma^{\alpha}}}, \:\text{as}\:\sigma\to0\\[3.0mm]
\displaystyle{\mathbb E T_{exit}(v_r \to v_l) \approx \frac{\alpha v_{r}^{\alpha}}{\sigma^{\alpha} }}, \:\text{as}\:\sigma\to0.
\end{array}\right.
\end{split}
\end{equation}
We note that the mean exit times in Eq.~\eqref{eq:sol1} depend on the location of the local minima $v_l$ and $v_r$. We further recall that the mean exit times of the processes driven by $\alpha$-stable noise are much shorter than those of Gaussian processes because of the presence of large jumps which occur with probability polynomially small in $\sigma$~\cite{Imkeller06}. 

On the other hand, the mean exit times of the Gaussian process follow Kramers' law~\cite{yamakou2018coherent,kramers1940brownian}, 
with escape events occurring with exponentially small probabilities, and are given by:
\begin{equation}\label{eq:sol2}%Eq. 18
\begin{split}
\left\{\begin{array}{lcl}
\displaystyle{\mathbb E T_{exit}(v_l \to v_r) \approx \exp\bigg(\frac{2\bigtriangleup U_{-}}{\sigma^2}\bigg)}, \:\text{as}\:\sigma\to0\\[4.0mm]
\displaystyle{\mathbb E T_{exit}(v_r \to v_l) \approx \exp\bigg(\frac{2\bigtriangleup U_{+}}{\sigma^2}\bigg), \:\text{as}\:\sigma\to0},
\end{array}\right.
\end{split}
\end{equation}
where $\bigtriangleup U_{\pm}$ are the energy barrier functions that depend, technically, on $w$ and $\phi$. 
The asymmetry of the potential in Eq.~\eqref{pot} is controlled only by the sign of the coefficient of the linear 
term, i.e., the sign of $w$. While the depths of the wells $\bigtriangleup U_{\pm}$ are controlled by the value of $w$ and more 
significantly, by the term $k_1\rho(\phi)/2$. But in the limit as $\varepsilon \to 0$ in Eq.~\eqref{eq:stoch}, the magnetic variable 
$\phi$ becomes almost constant and only the magnetic gain parameter $k_1$ now significantly changes the depths of the potential wells 
$\bigtriangleup U_{\pm}$. So we can drop the $\phi$ dependence in the energy barrier functions and write them as:
\begin{equation}\label{eq:sol3}
\begin{split}
\left\{\begin{array}{lcl}
\bigtriangleup U_{-}(w):=U_{k_1}(v_{m}) - U_{k_1}(v_{l}),\\[1.0mm]
\bigtriangleup U_{+}(w):=U_{k_1}(v_{m}) - U_{k_1}(v_{r}).
\end{array}\right.
\end{split}
\end{equation}
Thus, in the Gaussian case, the trajectories surmount the potential barriers $\bigtriangleup U_{\pm}$, such that the mean exit times depend exponentially on the depth of the potential well.  

We notice in Fig.~\ref{fig:3} that the depths of these barriers are inversely proportional to the strength of the magnetic gain parameter $k_1$. 
Thus, a stronger magnetic flux due to a larger value of $k_1$ should, on average, reduce the duration of the mean exit times of the trajectory perturbed 
by Gaussian noise, contributing to an increase in the spiking frequency. 

On the other hand, we also notice that the positions of the minima (at $v_l$ and $v_r$, indicated by the short vertical bars in Fig.~\ref{fig:3}) 
with respect to the fixed saddle (at $v_m=0.0$) change with $k_1$. We observe that the stronger magnetic flux  $k_1$, the smaller the distances of $v_l$ or $v_r$ 
from $0.0$, which in turn shortens, on average, the duration of the mean exit times of the trajectory perturbed by L\'{e}vy noise, contributing to an increase in the spiking frequency. 

From Eq.~\eqref{eq:1}, the deterministic timescale at which trajectories move on the stable parts of the 2-dimensional cubic nullcline of the current model, given by
$w(v,\phi) = -\frac{v^3}{3} + (1-k_1\rho(\phi))v$ (not shown), is $\varepsilon^{-1}$~\cite{yamakou2018coherent}. 
When there is no noise ($\sigma=0$), the neuron is in the excitable regime and as $\varepsilon\to0$, trajectories tend to spend a lot of time moving adiabatically 
along the stable parts of the 2D cubic nullcline, toward the unique stable fixed point at $(v_e,w_e,\phi_e)$ given by Eq.~\eqref{eq:11}, where it stops and stays 
for ever until a new perturbation is provoked by, e.g., a random process.

When noise is switched on ($\sigma\neq0$), it may kick a trajectory, which is moving quasi-deterministically at a timescale of $\varepsilon^{-1}$ along one stable 
branch of the 2D cubic nullcline, to another branch and then back. This corresponds to jumps out of the left and right potential wells, thereby causing a spike --- an oscillation. 
Depending on the type of noise perturbing the neuron, an escape from left to right (right to left) occurs at the stochastic timescale $\mathbb{E}T_{exit}$ given by the first (second) 
equation of Eq.~\eqref{eq:sol1} for the L\'{e}vy process or Eq.~\eqref{eq:sol2} for the Gaussian process.

It has been shown that the occurrence of SISR \textit{crucially} depends on the neuron's ability to asymptotically match, with probability close to \textit{unity}, 
the deterministic timescale $\varepsilon^{-1}$ (i.e., timescale at which a trajectory moves along the stable parts of the 2D cubic nullcline) and the stochastic 
timescale $\mathbb{E}T_{exit}$ (i.e., the timescale at which this trajectory escapes from the stable parts of this nullcline) at \textit{unique} 
exit points $w_-$ and $w_+$ located, respectively, on the left and right stable branches of the 2D cubic 
nullcline~\cite{yamakou2017simple,muratov2005self,deville2005two,muratov2008noise,deville2007nontrivial,deville2007self,shen2010self,yamakou2018coherent,yamakou2019control,10.3389/fncom.2020.00062}. 

If the deterministic timescale is shorter than the stochastic timescales (i.e., $\varepsilon^{-1}<\mathbb{E}T_{exit}$), 
then the trajectory has no time to escape from the left and right stable branches of the cubic nullcline which respectively 
correspond to the left and right wells of the potential $U_{k_1}(v)$. Because the neuron is in an excitable regime, 
the trajectory gets trapped in the left well of the potential (i.e., on the left stable branch of the cubic nullcline on which 
the unique stable fixed point is located) for too long. In this scenario, a spike is a rare event and this could destroy the coherence 
of the spiking, especially for short time intervals. 

On the other hand, if the deterministic timescale is longer than the stochastic timescales (i.e., $\varepsilon^{-1}>\mathbb{E}T_{exit}$), 
then the trajectory frequently escape from the potential wells (i.e., the stable branches of the cubic nullcline).  In this scenario, 
spiking is frequent (i.e., not rare) but incoherent because the trajectory escapes at several different points on the each of the stable branches of the cubic nullcline.

Interestingly, if at specific and unique points $w_-$ and $w_+$ on respectively the left and right stable branch of the cubic nullcline, 
the deterministic timescale matches the stochastic timescales (i.e., $\varepsilon^{-1}=\mathbb{E}T_{exit}$), \textit{frequent} and 
\textit{coherent} spiking emerges --- SISR occurs. The uniqueness of the exit points $w_-$ and $w_+$ can only be guaranteed by the 
\textit{monotonicity} of the minima $v_{l}(w)$ and $v_r(w)$ in the case of L\'{e}vy noise (see  Eq.~\eqref{eq:L1}) and the barrier 
functions $\bigtriangleup U_{-}(w)$ and $\bigtriangleup U_{+}(w)$ in the case of Gaussian noise (see Eq.~\eqref{eq:G1}). 

In Fig.~\ref{fig:4}, we show the graphs of the functions $|v_{l}|$, $v_r$, $\bigtriangleup U_{-}$, and $\bigtriangleup U_{+}$ 
with respect to $w\in[-\frac{2}{3},\frac{2}{3}]$, where the lower and upper bounds of this interval correspond to the $w$-coordinate of 
the local minimum and maximum of the cubic nullcline, respectively. Here, we see that these functions are all monotone with respect 
to $w\in[-\frac{2}{3},\frac{2}{3}]$. Hence, frequent and coherent spiking would occur if we match the deterministic and stochastic 
timescales only at $w_-$ on the left stable branch and at $w_+$ on the right stable branch of the cubic nullcline, that is:
\begin{figure}%[htb!]
\begin{center}
\includegraphics[width=9.0cm,height=5.0cm]{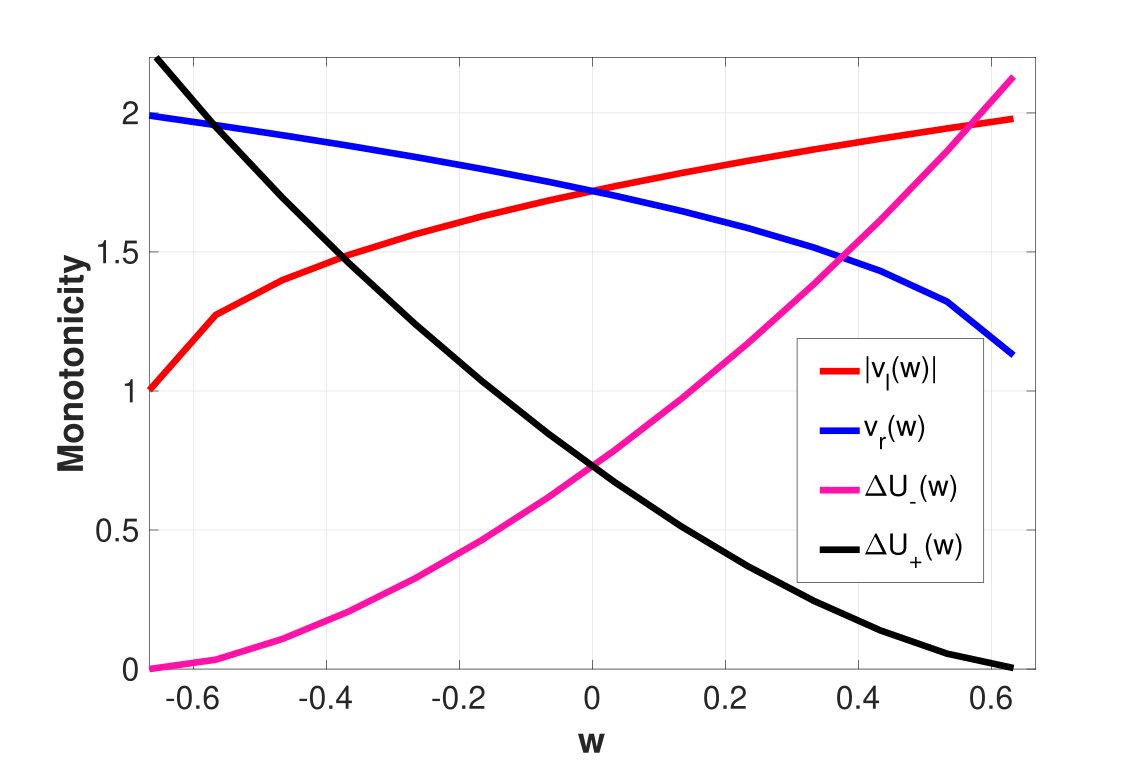}
\end{center}
\caption{{The graphs of the $|v_{l}(w)|$, $v_{r}(w)$, $ \bigtriangleup U_{-}(w)$, and $\bigtriangleup U_{+}(w)$ with  respect to $w\in[-\frac{2}{3},\frac{2}{3}]$. Their monotonicity ensure the uniqueness of the escape points $w_-$ and $w_+$ which satisfy the equations in Eq.~\eqref{eq:L1} and Eq.~\eqref{eq:G1}. 
Parameters are $k_1=0.1$ and $\phi=0.85$.}}\label{fig:4}
\end{figure}
\begin{eqnarray}\label{eq:L1}
\begin{split}
\displaystyle{\frac{\alpha|v_{l}(w_-)|^{\alpha}}{\sigma^{\alpha}}}=\frac{1}{\varepsilon}=
\displaystyle{\frac{\alpha v_{r}(w_+)^{\alpha}}{\sigma^{\alpha} }},
\end{split}
\end{eqnarray}
for the L\'{e}vy process, and
\begin{eqnarray}\label{eq:G1}
\begin{split}
 \displaystyle{\exp\bigg(\frac{2\bigtriangleup U_{-}(w_-)}{\sigma^2}\bigg)}=\frac{1}{\varepsilon}=
 \displaystyle{\exp\bigg(\frac{2\bigtriangleup U_{+}(w_+)}{\sigma^2}\bigg)},
\end{split}
\end{eqnarray}
for the Gaussian process. Therefore, the occurrence of SISR (i.e., \textit{frequent} and \textit{coherent} spiking activity) will
depend on the neurons' ability to asymptotically match the timescales by taking the following double scaling limits:
\begin{eqnarray}\label{eq:L2}
\begin{split}
\displaystyle{\lim \limits_{(\varepsilon,\sigma)\rightarrow (0,0)}}\bigg[\sigma^{\alpha}\varepsilon^{-1}\bigg]\to
\left\{\begin{array}{lcl}
\displaystyle{\alpha|v_{l}(w_-)|^{\alpha}}\\[2.0mm]
\displaystyle{\alpha v_{r}(w_+)^{\alpha}}
\end{array}\right.
\end{split}
\end{eqnarray}
for the L\'{e}vy process, and
\begin{eqnarray}\label{eq:G2}
\begin{split}
\displaystyle{\lim \limits_{(\varepsilon,\sigma)\rightarrow (0,0)}}\Bigg[\frac{\sigma^2\ln(\varepsilon^{-1})}{2}\Bigg]\to
\left\{\begin{array}{lcl}
 \displaystyle{\bigtriangleup U_{-}(w_-)}\\[2.0mm]
 \displaystyle{\bigtriangleup U_{+}(w_+)}
\end{array}\right.
\end{split}
\end{eqnarray}
for the Gaussian process~\cite{muratov2005self,yamakou2018coherent}.

Due to the anomalous long jumps of a trajectory perturbed by a L\'{e}vy process~\cite{Imkeller06,koren2007leapover,dybiec2016hit,ditlevsen1999anomalous}, 
this trajectory does not necessarily have to hit the saddle point at $v_m$ before escaping from the stable branches of the 2D cubic nullcline. Hence, escapes 
may instantaneously occur even with a very weak noise intensity. This means that the ``frequent spiking'' requirement of SISR can be easily achieved by a 
L\'{e}vy process, even with a very weak intensity. However, the ``coherent spiking'' requirement of SISR can only be guaranteed by the asymptotic scaling limits given in Eq.~\eqref{eq:L2}.  

In the Gaussian case, a trajectory can only escape from a potential well after hitting the boundary at the saddle point at $v_m$. Therefore, 
the ``frequent spiking'' requirement of SISR needs that the noise intensity is not \textit{too} weak (otherwise, we get a Poissonian spike train --- a rare spiking 
event which could destroy the coherence of the spiking~\cite{yamakou2018coherent}). Moreover, we observe that the stochastic timescales of the Gaussian noise 
in Eq.~\eqref{eq:sol2} depend on the energy barrier functions $\bigtriangleup U_{\pm}$. If these barriers are too deep (i.e., $\bigtriangleup U_{\pm} \to \infty$), 
then weak noise intensities cannot provoke escapes (at least frequently), and the trajectory will remain strapped inside a potential well. Thus, the noise has be to
weak (so that the mean exit times satisfy Eq.~\eqref{eq:sol2}), but strong enough to able to invoke some spiking. If this Gaussian noise is strong enough to invoke 
spiking, then the ``coherent spiking'' requirement of SISR can only be guaranteed by the asymptotic scaling limits given by Eq.~\eqref{eq:G2}.  Thus, for L\'{e}vy 
noise, we expect SISR to occur even at very weak noise intensities. But for Gaussian noise, we expect SISR to occur at a comparatively larger intensity. 

To answer the three main questions we are interested in (see the introduction section), we will set the memristive neuron in the excitable regime by choosing $c=0.95$, $a=0.1$, $b=0.02$, $d=0.5$, 
and also set location parameter of the standardized L\'{e}vy process at $\mu=0.0$. We chose a sufficiently small timescale 
separation parameter, i.e., $\varepsilon=0.001\ll1$, weak noise intensity, i.e., $0<\sigma<1$, and then numerically search for the combined values 
of $k_1\in[0.0,2.0]$, $k_2\in[1.0,2.0]$, $\alpha\in(0,2]$, and $\beta\in[-1,1]$ for which the scaling limit conditions in Eq.~\eqref{eq:L2} 
and Eq.~\eqref{eq:G2} are satisfied (or at least to some degree) or not.

\section{Numerical results and discussion}\label{Sec. IV}
To measure the degree of SISR (i.e., the degree to which Eq.~\eqref{eq:L2} 
and Eq.~\eqref{eq:G2} are satisfied), we use the coefficient of variation ($CV$), an important statistical measure based on the time intervals 
between spikes~\cite{pikovsky1997coherence}. From a neurobiological point of view, $CV$ is more important than other measures 
(e.g., power spectral density and auto-correlation function) because it is related to the timing precision of information processing in neural systems~\cite{pei1996noise}.
$CV$ uses the inter-spike intervals (ISIs) where the $k$th interval is 
the  difference between two consecutive spike times $t^k$ and $t^{k+1}$ of the neuron, and is defined as:
\begin{equation}\label{eq:cv}
CV=\dfrac{\sqrt{\langle ISI^2\rangle-\langle ISI\rangle^2}}{\langle
ISI\rangle},
\end{equation}
where $\langle ISI\rangle$ and $\langle ISI^2\rangle$ represent
the mean and the mean squared ISIs, respectively.
When $CV=1$, we have Poissonian spike train (i.e., rare and incoherent spiking), 
and when $CV>1$ we have a point process that is even more variable than a Poisson process~\cite{kurrer1995noise}. In both these cases, 
the degree of SISR is quite low as the double limits in the left-hand sides of Eq.~\eqref{eq:L2} and Eq.~\eqref{eq:G2} fail to converge toward the
corresponding values on the right-hand sides. The degree of SISR  becomes higher with $CV \to 0$ as the double limits in the left-hand sides of 
Eq.~\eqref{eq:L2} and Eq.~\eqref{eq:G2} also converge toward the corresponding values on the right-hand sides. When $CV=0$, the double limits in 
the left-hand sides of Eq.~\eqref{eq:L2} and Eq.~\eqref{eq:G2} should be exactly equal to the corresponding values on the right-hand sides. 
In this case, we will have perfectly  ``deterministic'' periodic  spiking.

For our numerical simulations, we used the fourth-order stochastic Runge-Kutta algorithm employed in \cite{huang2011effects,xu2013levy,wang2016levy} and proven in \cite{ruemelin1982} to strongly converge. In should be noted that for general noise, the numerical solution of stochastic differential equations that uses the scheme proposed by Wilkie \cite{wilkie2004} may not be intact even with additive noise, see also \cite{burrage2006}.

We generate the L\'{e}vy random variable by using the Janicki-Weron algorithm~\cite{janicki1993simulation} which has been proven \cite{zolotarev1983one,weron1996chambers} to generate stable random variable for all admissible values of the parameters $\alpha$, $\beta$ $\mu$, and $\sigma$.
%It is worth noting that the characteristic
%functions of the $\alpha$-stable distribution are not continuous functions of the parameters determining %them, they have discontinuities at all points of the form $\alpha = 1$; $\beta \neq0$.
%Hence, apart from a few exceptions (Gaussian, Cauchy, and Lévy-Smirnoff, for which simple methods of simulation have been found), there are no analytic expressions for the inverse of the stable distribution function and the inverse transform method cannot be used. However, the Janicki-Weron algorithm (used in our simulations) has been proven \cite{zolotarev1983one,weron1996chambers} to generate stable random variable for all admissible values of the parameters $\alpha$, $\beta$ $\mu$, and $\sigma$.
We numerically integrate Eq.~\eqref{eq:stoch} for a very long time interval (i.e., {$T=4\times10^{7}$} time unit which allows for the small value of $\varepsilon=0.001$, the  collection of sufficiently many ISIs for statistical estimate). We then average the ISIs over time and up to {$30$} realizations for each noise amplitude.

We recall that the continuous jump property of a Gaussian process (with finite variance) forces the trajectories to hit the boundary of a domain before escaping. While with the discontinuous long-jumps of a L\'{e}vy process with $\alpha<2$ (with infinite variance), trajectories can rapidly escape to infinity without hitting the boundary. Thus, for our FHN neuron perturbed by a L\'{e}vy noise, we might need to wait for a long time for a trajectory which had exhibited a long-jump to come back to the vicinity of the stable fixed point, if there is no compulsory truncation. It is important to note that these long waiting times can significantly affect the ISIs. Hence, because the $CV$ (used to characterised the degree of SISR) depends (only) on the ISIs, the numerical results obtained would be sensitive to the choice of the truncation threshold. Considering the physical and computer saturation effects, a suitable truncation scheme should, therefore, be adoptable.  In our simulations, we use the truncation threshold $v=3.0$ $\times$ sign$(v)$ whenever $\lvert v\rvert>3.0$. This is a well-known truncation scheme for $\alpha$-stable noises employed in many relevant references \cite{kosko2001robust,mitaim2004adaptive,liu2018stochastic}.

\begin{figure}%[htb!]
\begin{center}
\includegraphics[width=8.0cm,height=5.0cm]{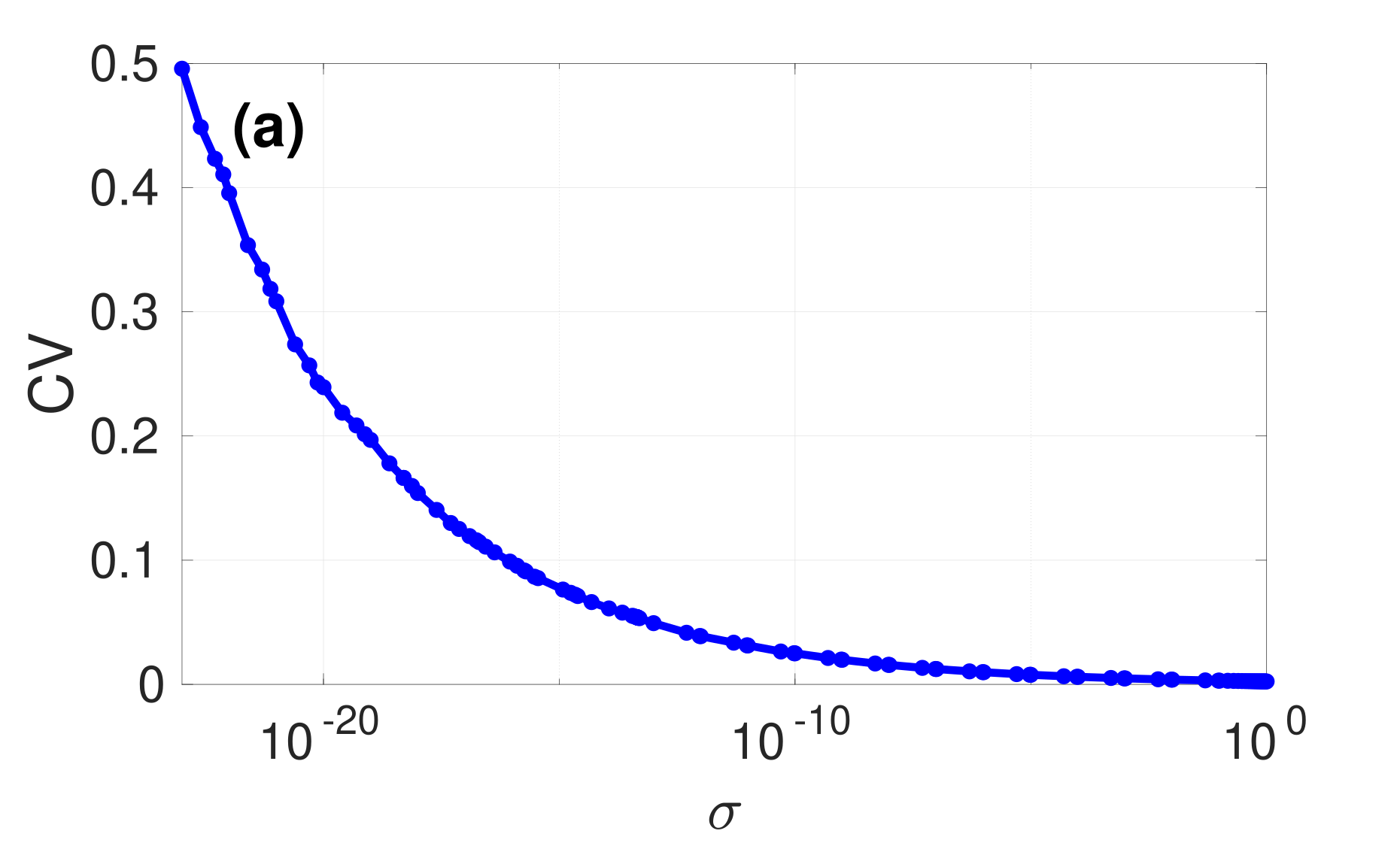}\includegraphics[width=8.0cm,height=5.0cm]{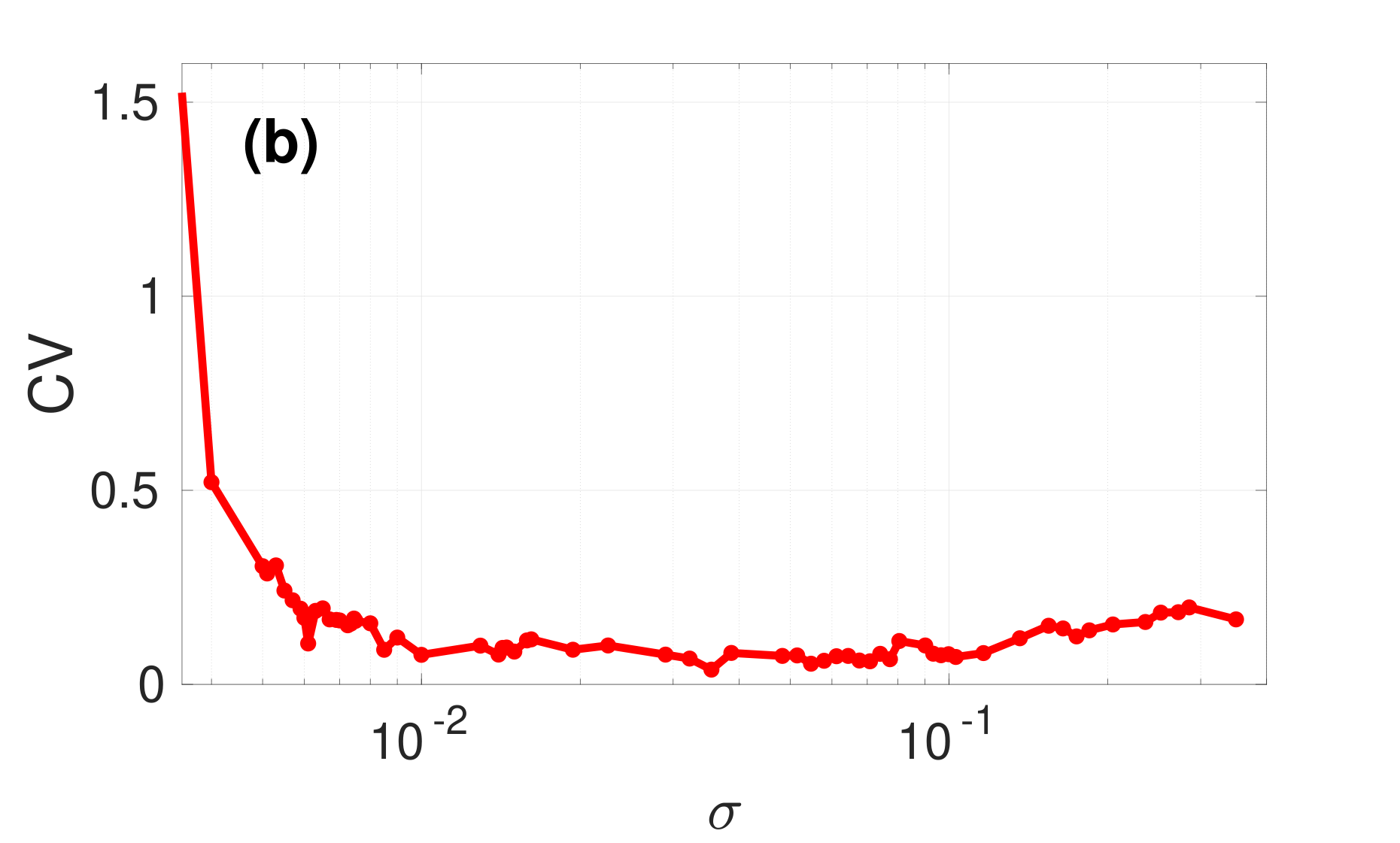}
\includegraphics[width=8.0cm,height=5.0cm]{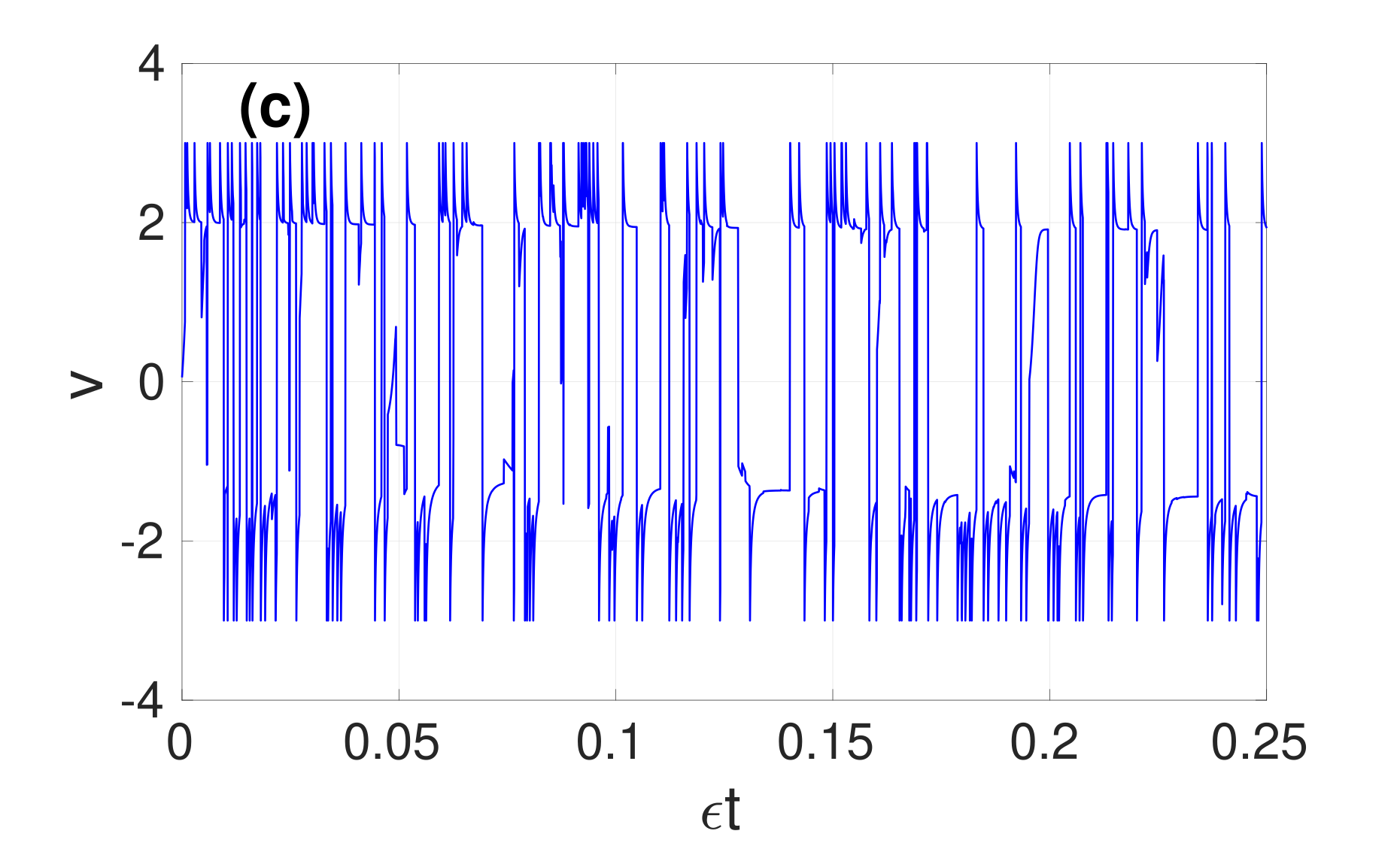}\includegraphics[width=8.0cm,height=5.0cm]{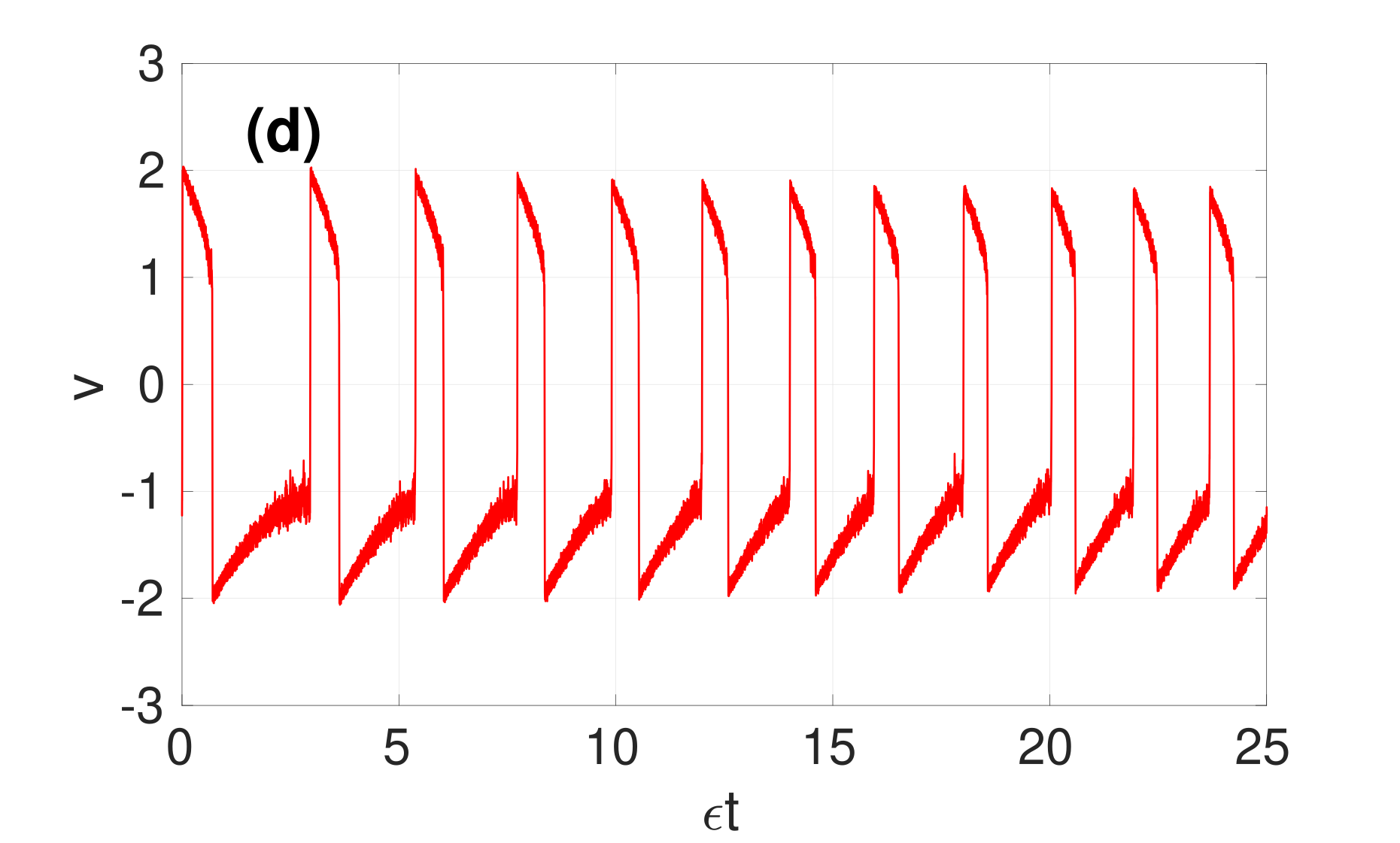}
\end{center}
\caption{{The variation of $CV$ with the noise intensity $\sigma$ with L\'{e}vy noise $(\alpha=0.1$, $\beta=0.0)$ in \textbf{(a)} and Gaussian noise $(\alpha=2.0$, $\beta=0.0)$ in \textbf{(b)}.
Time series during SISR induced by the L\'{e}vy noise in \textbf{(c)} with $\sigma=0.04$ and Gaussian noise in \textbf{(d)} with $\sigma= 0.04$. Degree of SISR is higher with L\'{e}vy noise than with Gaussian noise for all values of $\sigma$. $k_1=0.1$ and $k_2=0.1$.}}\label{fig:5}
\end{figure}

To avoid the long waiting times to which CV is sensitive to, we decided to use the truncation threshold above. We note that the threshold values (i.e., $v=-3$ and $v=3$) are respectively below and above, but also sufficiently close to the extreme values ($v=-2$ and $v=2$, see Fig.~\ref{fig:5}\textbf{(d)}) of the \textit{relaxation oscillations} of the underlining deterministic FHN model. A value of, for example, $v=1000$ is not physiological for the FHN model. Thus, the truncation threshold used not only ensures that the simulated trajectories do not escape to infinity (thereby avoiding the long waiting times) but also ensures that the trajectories go not too far below and above the extreme values of the relaxation oscillation, which are in fact the physiologically acceptable extreme values for the model. In the presence of noise, the random trajectories may then oscillate with slightly bigger amplitudes compared to that of the deterministic relaxation oscillation.  Thus, the truncation scheme used gives room for these fluctuations to be taken into account without any significant effect on the waiting times that arise due to the long-jumps. These makes the truncation threshold $v = 3 \times sign(v)$ whenever $|v|>3$, a good ans natural choice when calculating the CV values of the FHN model perturbed by a L\'{e}vy noise.

Fig.~\ref{fig:5}\textbf{(a)} and \textbf{(c)} respectively show the variation of $CV$ with the noise intensity $\sigma$ for a very impulsive ($\alpha=0.1$) and symmetric ($\beta=0.0$) L\'{e}vy noise and a time series of the coherent spike trains obtained at a noise intensity that satisfies Eq.~\eqref{eq:L2}. 
The $CV$-curve and time series are computed in a weak magnetic flux regime ($k_1=0.1$, $k_2=0.1$) and show that as long 
as Eq.~\eqref{eq:L2} is valid, L\'{e}vy noise can $(i)$ induce a high degree of SISR even at very weak noise intensities  (e.g., {$CV\approx0.075$ at $\sigma=1.0\times10^{-15}$)}, and  $(ii)$ induce an even higher degree of SISR at relatively larger noise intensities (e.g., {$CV=0.0015$ at $\sigma\approx0.9$).} It is worth noting that in Fig.~\ref{fig:5}\textbf{(a)} and \textbf{(c)} the L\'{e}vy noise is very impulsive, i.e., the stability index is very small ($\alpha=0.1$), and therefore even at very weak noise intensities (such as $\sigma=1.0\times10^{-20}$),
the long-jumps can still occasionally occur, thereby inducing some spikes whose $ISIs\neq0$ will contribute to a finite CV value. But as $\alpha$ increases, the long-jumps become less frequent and of shorter range. Thus, only relatively larger noise intensities can invoke spikes as in Gaussian case in
Fig.~\ref{fig:5}\textbf{(b)} and \textbf{(d)}. 

In Fig.~\ref{fig:5}\textbf{(b)} and \textbf{(d)}, we respectively show the variation of $CV$ with the noise intensity $\sigma$ for Gaussian noise ($\alpha=2.0$, $\beta=0.0$) and a time series of the coherent spike train obtained at a noise 
intensity which satisfies Eq.~\eqref{eq:G2}, in the same weak magnetic flux regime ($k_1=0.1$, $k_2=0.1$). Comparing the degree 
of SISR induced by a L\'{e}vy noise with parameters at $\alpha=0.1$ and $\beta=0.0$ to that of Gaussian noise ($\alpha=2.0$, $\beta=0.0$), 
we see that L\'{e}vy noise can induce a higher degree of SISR with both extremely weak and weak noise amplitudes. In Fig.~\ref{fig:5}\textbf{(b)} 
with Gaussian noise, we have a low (and almost constant) {$CV\approx0.045$ \textit{only} in the weak (but not too weak) noise intensities, i.e., for $\sigma\in( 0.01,0.1)$.}

\begin{figure}%[htb!]
\begin{center}
\includegraphics[width=8.0cm,height=5.0cm]{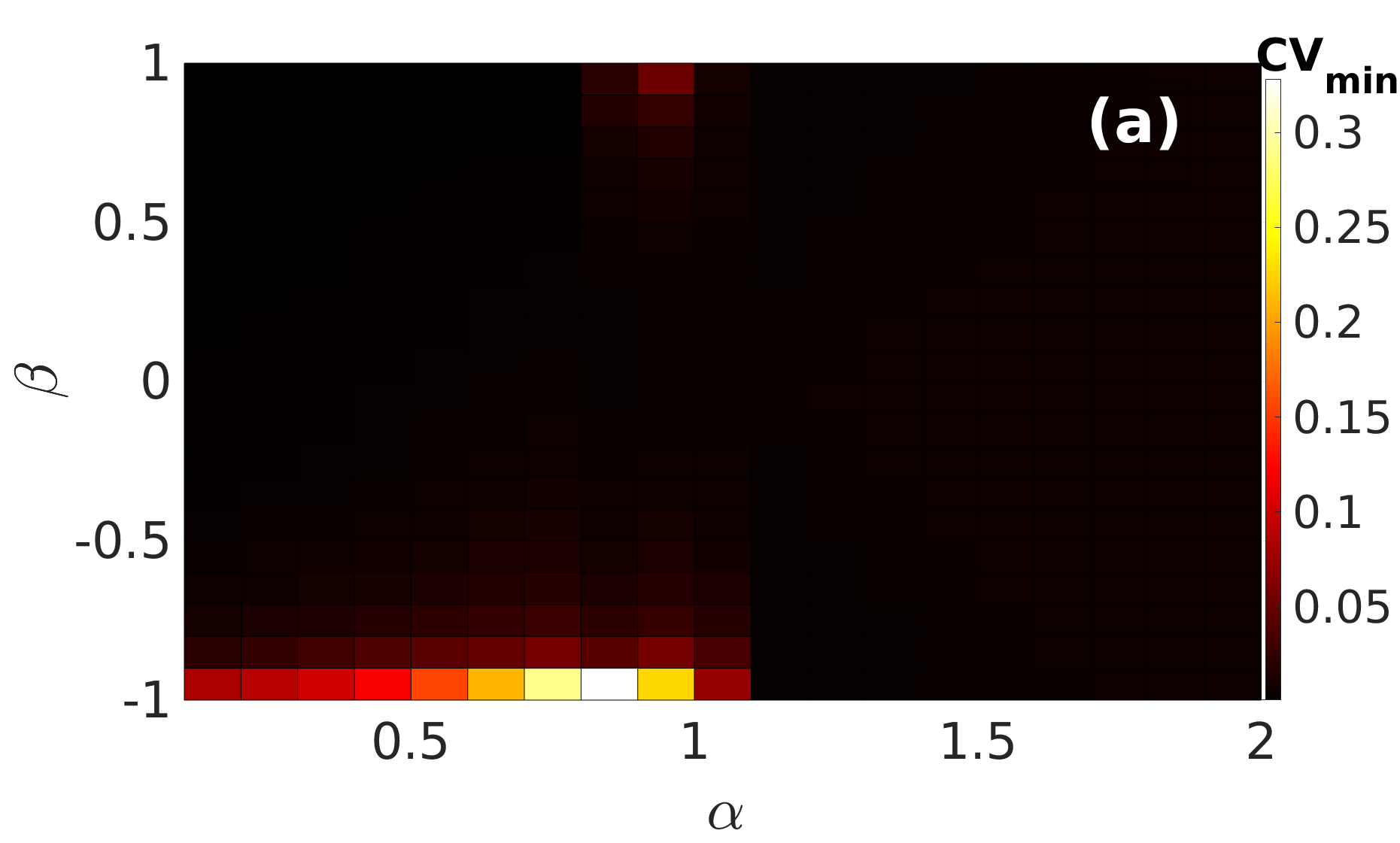}\includegraphics[width=8.0cm,height=5.0cm]{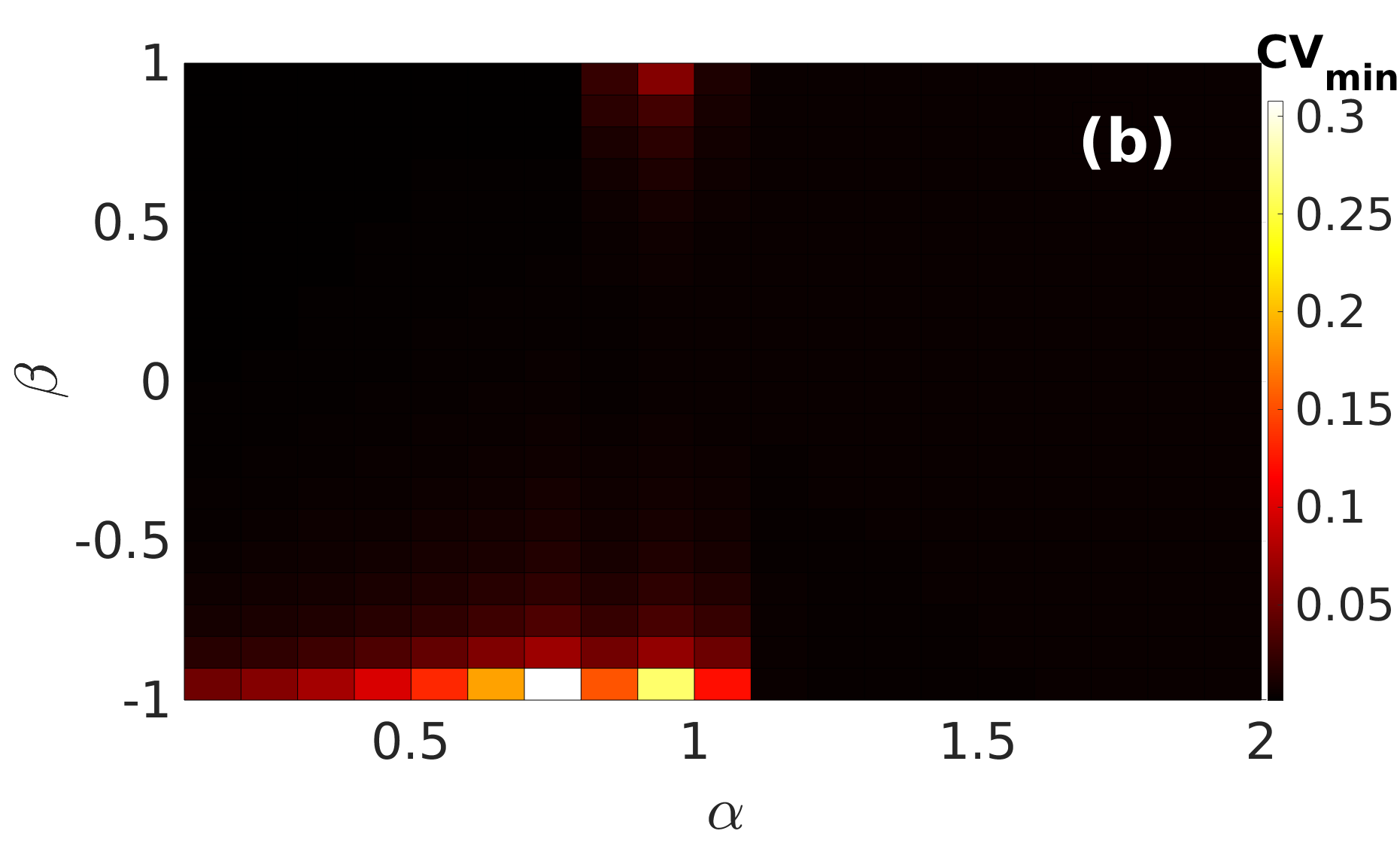}
\end{center}
\caption{{Variations of the minimum CV ($CV_{min}$) with respect to the stability index ($\alpha$) and the skewness ($\beta$) parameters with weak ($k_1=0.1$, $k_2=0.1$) and strong ($k_1=2.0$, $k_2=1.0$) magnetic gain parameters in \textbf{(a)} and \textbf{(b)}, respectively.}}\label{fig:6}
\end{figure}

Fig.~\ref{fig:6}\textbf{(a)} and \textbf{(b)} show minimum coefficient of variation ($CV_{min}$) against the stability index ($\alpha$) and the 
skewness ($\beta$) parameters of the L\'{e}vy process in a weak ($k_1=0.1$ and $k_2=0.1$) and in a strong ($k_1=2.0$ and $k_2=1.0$) magnetic flux regime, respectively. 

In Fig.~\ref{fig:6}\textbf{(a)}, with a weak magnetic flux regime ($k_1=0.1$, $k_2=0.1$), a right-skewed (i.e., {$\beta\in(0.0,1.0]$) L\'{e}vy process with a low stability index (i.e., $\alpha\in(0.0,0.7]$}) can induce a high degree of SISR, as indicated by the very low value of {$CV_{min}\approx0.0014$.} 
With higher values of $\alpha$, i.e., for $\alpha\in(1.0,2.0)$ and irrespective of the value of the skewness parameter, 
i.e., for $\beta\in[-1.0,1.0]$, the degree of SISR is high and almost constant as indicated by the low and almost constant 
{$CV_{min}\approx0.005$.} Even though this cannot be clearly seen from the panel, the data shows that, at $\alpha=2.0$ and $\beta\in[-1.0,1.0]$ (which includes the Gaussian case at $\beta=0.0$), the $CV_{min}$ is also the low and almost constant at {$CV_{min}\approx0.0497$}, i.e, almost 10 order of magnitude higher than the $CV_{min}$ of the L\'{e}vy processes in which {$\alpha\in(1.0,2)$ and $\beta\in[-1.0,1.0]$.} And for {$\alpha\in(0.0,1.0]$ and $\beta\in[-1.0,-0.5]$} (i.e., from the bright red, the yellow, and the white regions), the degree of SISR is relatively low, as $CV_{min}$ continuously vary in the 
interval {$CV_{min}\in[0.125, 0.328]$} with the highest value at {$CV_{min}\approx0.328$,} occurring at $\alpha=0.8$ and $\beta=-1.0$.

In Fig.~\ref{fig:6}\textbf{(b)}, with a strong magnetic flux regime ($k_1=2.0$, $k_2=1.0$), the variation in the degree of SISR is 
qualitatively the same as in  Fig.~\ref{fig:6}\textbf{(a)}, but data show that there is a slight quantitative difference in the oder of magnitude of the $CV_{min}$ values, and hence in the degree of SISR in both panels. 
For example, when we have Gaussian noise (i.e., $\alpha=2.0$ and $\beta=0.0$), we have a {$CV_{min}\approx0.0538$} for weak magnetic flux in Fig.~\ref{fig:6}\textbf{(a)} and {$CV_{min}\approx0.0497$} for strong magnetic flux in Fig.~\ref{fig:6}\textbf{(b)}. Later, we shall discuss and show more clearly in the $(k_1,k_2)$-plane the effects of the magnetic gain parameters on the degree of SISR.

The presence of intermittent intervals of sub-threshold spiking explains the relatively high values of {$CV_{min}\in[0.125, 0.328]$ in the region
bounded by $\alpha\in(0.0,1.0]$ and $\beta\in[-1.0,-0.5]$} (i.e., the bright red, yellow, and white regions) in the panels of Fig.~\ref{fig:6}. Because of these intervals of intermittent sub-threshold spiking (with $v\leq v_{th}=1.3$, an arbitrarily chosen threshold value), the regularity of the ISIs which is calculated based on the occurrence of supra-threshold spiking (with $v> v_{th}$) is deteriorated. On the other hand, for parameter values  in the regions bounded by {$\alpha\in(0.0,1.0]$ and $\beta\in(-0.5,1.0]$ (i.e., dark region with $CV_{min}\approx0.0014$), $\alpha\in(1.0,2.0)$ and $\beta\in[-1.0,1.0]$ (i.e., dark region with $CV_{min}\approx0.005$), and by $\alpha=2.0$ and $\beta\in[-1.0,1.0]$ (i.e., dark region with $CV_{min}\approx0.0497$),} the time series contain fewer intermittent intervals of sub-threshold spiking (see, e.g., Fig.~\ref{fig:5}\textbf{(c)}), hence the low value of the $CV_{min}$ in these regions. 

\begin{figure}%[htb!]
\begin{center}
\includegraphics[width=5.5cm,height=5.0cm]{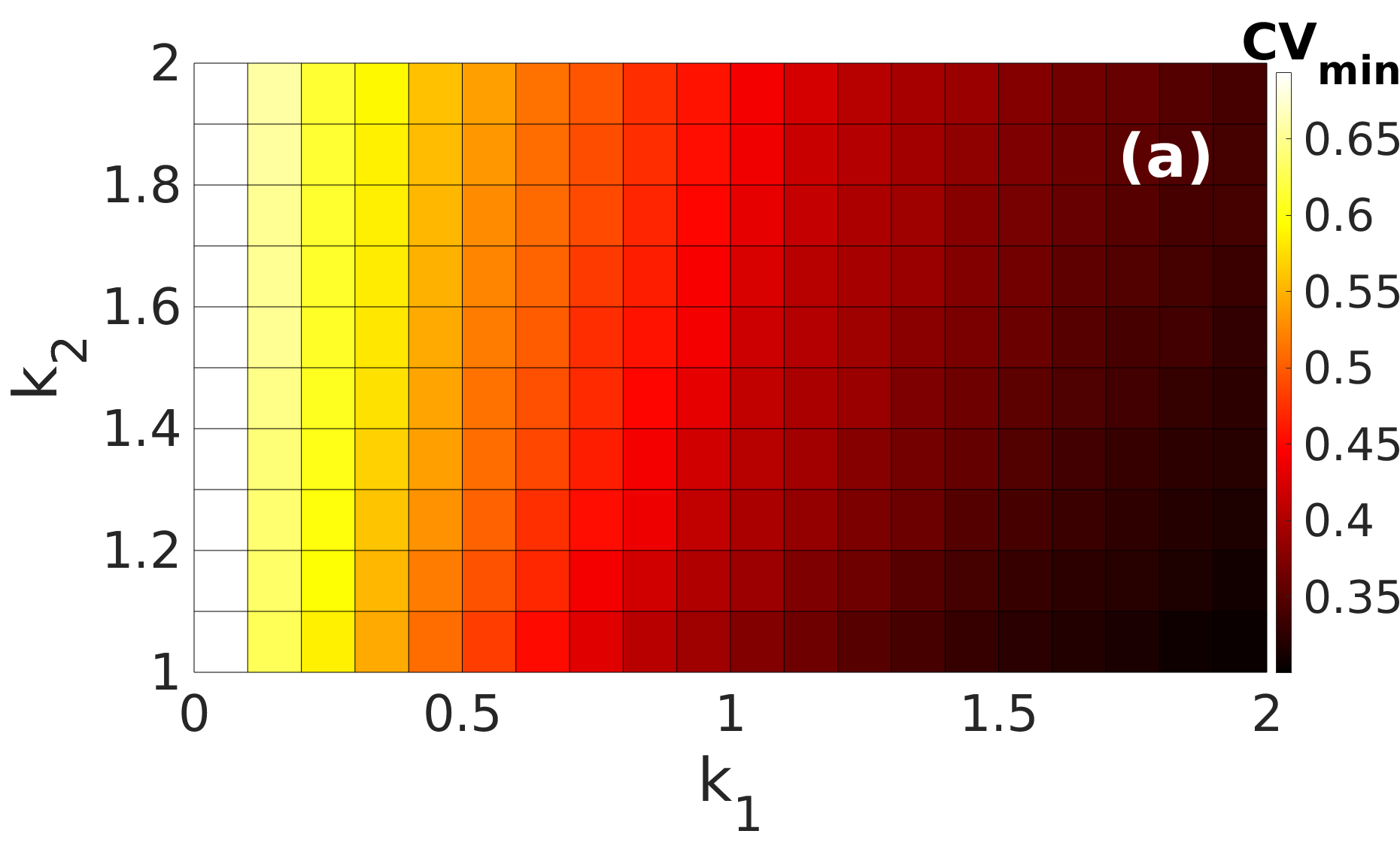}\includegraphics[width=5.5cm,height=5.0cm]{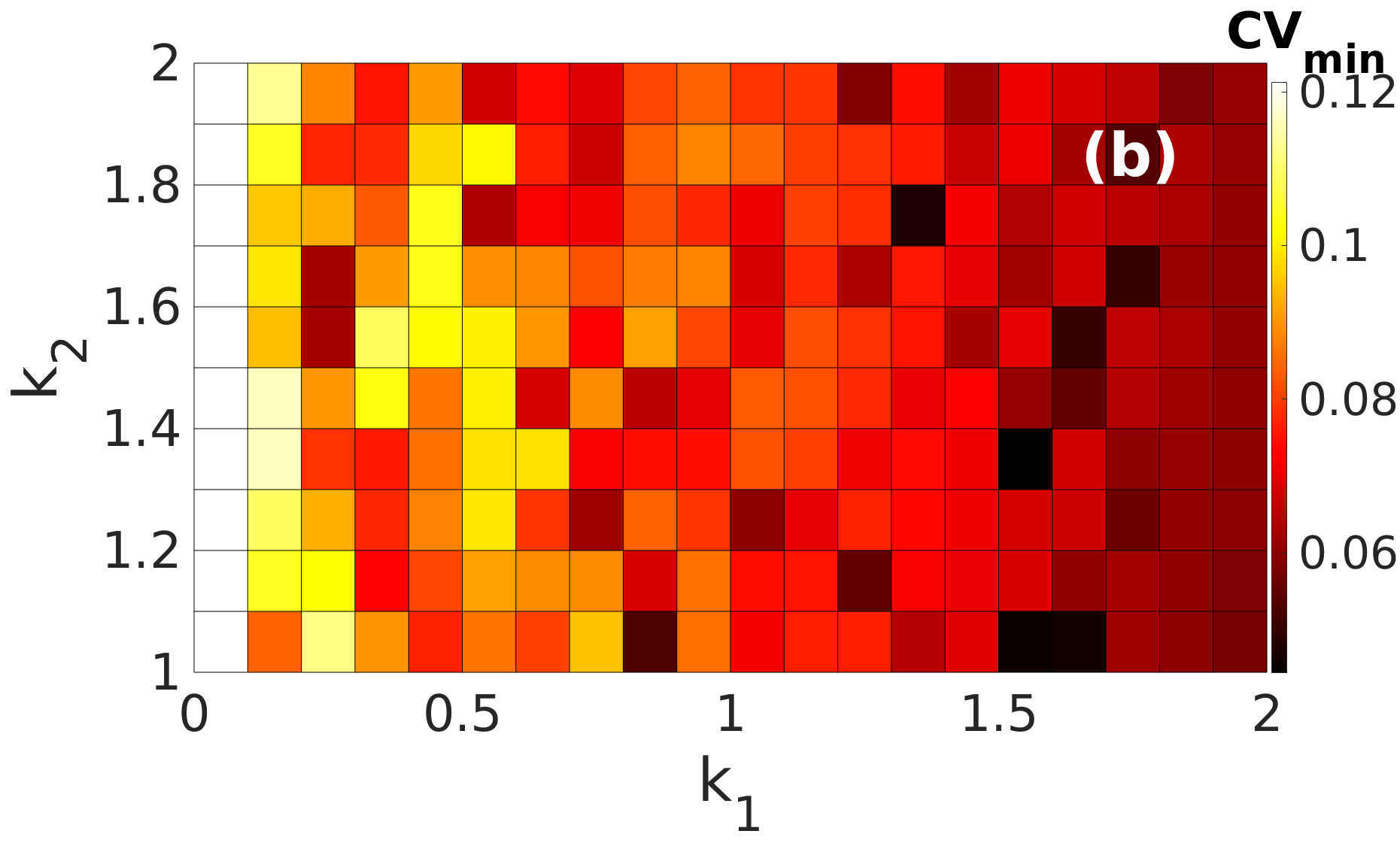}\includegraphics[width=5.5cm,height=5.0cm]{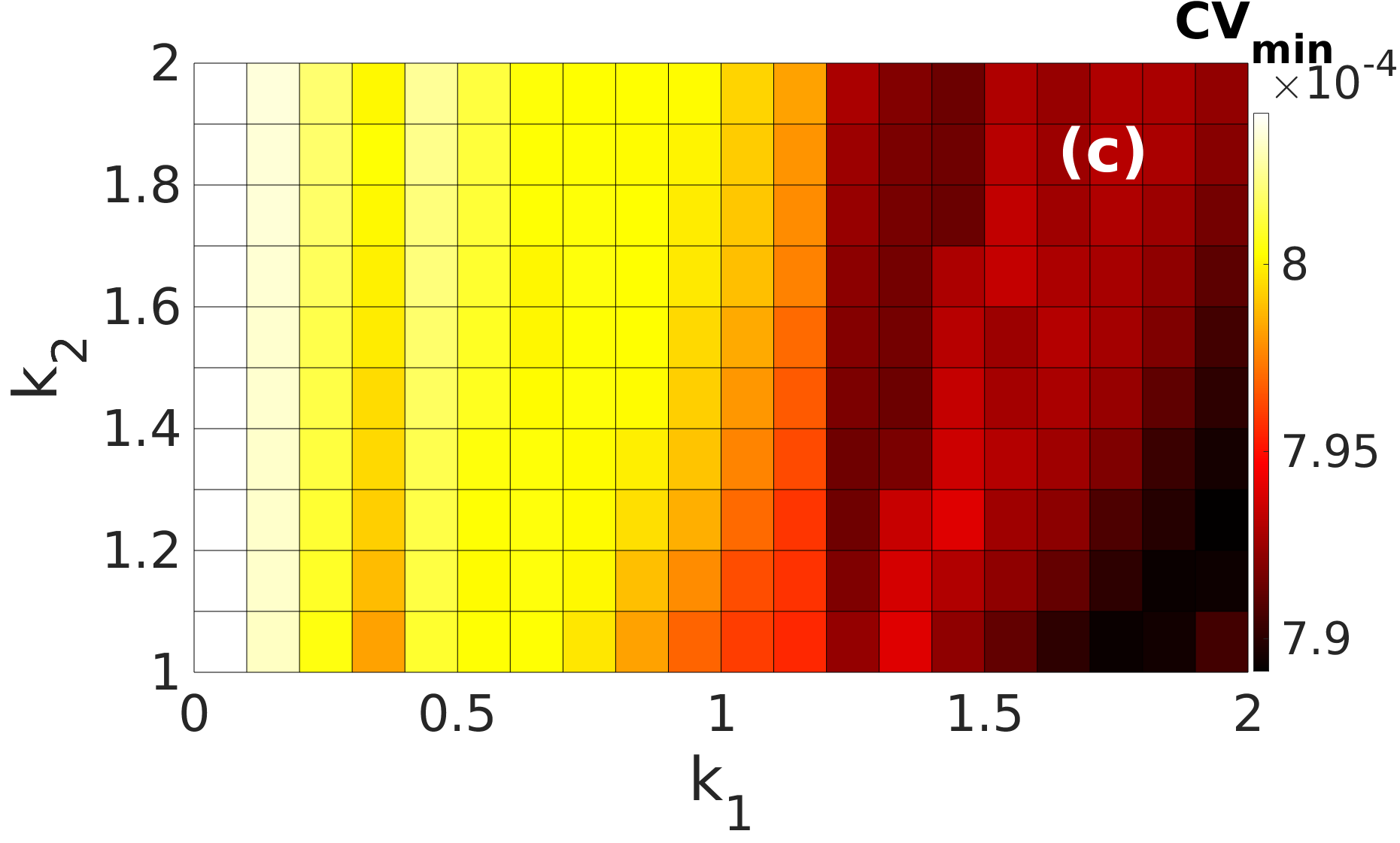}
\end{center}
\caption{{Variations of the minimum CV ($CV_{min}$) with respect to the magnetic gain parameters $k_1$ and $k_2$ at different values of the stability index and skewness parameters. In all cases, the larger $k_1$  is and the smaller $k_2$ is, the lower is the value of $CV_{min}$, i.e., the higher the degree of SISR.
In \textbf{(a)}: $\alpha=0.7$, $\beta=-1.0$; in \textbf{(b)}: $\alpha=2.0$,  $\beta=0.0$; and \textbf{(c)}: $\alpha=0.1$, $\beta=1.0$.}}\label{fig:7}
\end{figure}

In Fig.~\ref{fig:7}, we show the variation in the degree of SISR with the variations in the strengths of the magnetic gain parameters
$k_1$ and $k_2$ in three specific regions of interest in Fig.~\ref{fig:6}\textbf{(a)}: (i) when the degree of SISR is 
low, i.e., in the white spot with $\alpha=0.7$ and $\beta=-1.0$, (ii) when the degree of SISR is high, i.e., the dark red 
region with $\alpha=2.0$ and $\beta=0.0$ (i.e., Gaussian), and (iii) when the degree of SISR is very high, i.e., the black 
region with $\alpha=0.1$ and $\beta=1.0$. We also note that in all the panels of Fig.~\ref{fig:7}, the magnetic gain parameter 
$k_2$ is restricted to $k_2\geq1.0$, so that the memristive neuron always lies in the excitability region (black region) for all 
values of $k_1\geq0.0$, as indicated in Fig.~\ref{fig:2}\textbf{(b)}.

In Fig.~\ref{fig:7}\textbf{(a)}, we can now clearly see the effects of the magnetic gain parameters on the degree of SISR when
$\alpha=0.7$ and $\beta=-1.0$, corresponding, from Fig.~\ref{fig:6}\textbf{(a)}, to the white spot with a relatively large { $CV_{min}\approx0.328$.}
We observe that: the \textit{stronger} the magnetic gain parameter $k_1$ --- that bridges the coupling and modulation on the 
membrane potential $v$ from magnetic field $\phi$ --- and the \textit{weaker} the parameter $k_2$ --- that describes the degree 
of polarization and magnetization by adjusting the saturation of magnetic flux --- the higher the degree of SISR. 
In Fig.~\ref{fig:7}\textbf{(a)}, as $k_1\rightarrow2.0$ and $k_2\rightarrow1.0$, the color-coded $CV_{min}$ goes from a white 
region with a relatively high value of {$CV_{min}\approx0.69$}, via a yellow and a red, to a black region with the lowest {$CV_{min}\approx0.30$.}
Moreover, irrespective of the value of $k_2$,
when $k_1=0.0$, $CV_{min}$ takes the highest value of the panel (i.e., {$CV_{min}\approx0.69$} in the white region). 
Further numerical simulations (not shown) indicated that this behavior is qualitatively the same for many pairs of values of {$\alpha\in(0.0,1.0]$ and $\beta\in[-1.0,-0.5]$.} 
This means that the appropriate combination of  values of the magnetic gain parameters can significantly improve the degree of SISR 
induced by L\'{e}vy noise when the noise parameters are in intervals {$\alpha\in[0.0,1.0]$ and $\beta\in[-1.0,-0.5]$.} We shall see 
later in Fig.~\ref{fig:7}\textbf{(c)} that this significant improvement in the degree of SISR depends on intervals in which $\alpha$ and $\beta$ are located.

In Fig.~\ref{fig:7}\textbf{(b)}, we have Gaussian noise (i.e., $\alpha=2.0$ and $\beta=0.0$) and effects of the magnetic gain parameters 
are qualitatively the same as in Fig.~\ref{fig:7}\textbf{(a)} with a L\'{e}vy noise having parameters at $\alpha=0.7$ and $\beta=-1.0$. 
That is, the weaker $k_2$ and the stronger $k_1$ become, the \textit{lower} is $CV_{min}$, on average. 

It is worth noting, by comparing
Fig.~\ref{fig:7}\textbf{(a)} and \textbf{(b)}, that the degree of SISR induced by L\'{e}vy noise (with $\alpha=0.7$ and $\beta=-1.0$) is lower than that induced by Gaussian noise ($\alpha=2.0$ and $\beta=0.0$). Furthermore, the effects of the magnetic gain parameters $k_1$ and $k_2$ on the degree of SISR is weaker in the Gaussian case. 
That is, in Fig.~\ref{fig:7}\textbf{(b)}, $CV_{min}$ varies in the interval $[0.044,0.121]$, compared to $[0.30,0.69]$ in Fig.~\ref{fig:7}\textbf{(a)}. The bigger range in the latter interval
indicates the stronger effects of the magnetic gain parameters on the degree of SISR induced by L\'{e}vy noise when its parameters lie in the intervals {$\alpha\in(0.0,1.0]$ and $\beta\in[-1.0,-0.5]$.}

Moreover, it important to note that the degree of SISR in the non-memristive neuron (i.e., when $k_1=0$) is always lower (poorer) than that in the memristive one. This result is confirmed by comparing $CV_{min}$ in the non-memristive FHN neuron perturbed by Gaussian noise --- studied in our previous work~\cite{yamakou2018coherent} --- to the memristive FHN model studied in the current paper. In the non-memristive case, the 
lowest $CV$ value is always at $CV\approx0.2$, while in the memristive case, the lowest value gets even smaller, {i.e., $CV\approx0.044$,}
especially as $k_1 \to 2$ and $k_2 \to 1$.

In Fig.~\ref{fig:7}\textbf{(c)}, we have a L\'{e}vy noise with $\alpha=0.1$ and $\beta=1.0$, which corresponds to a black region
(i.e., with a high degree of SISR) in Fig.~\ref{fig:6}\textbf{(a)}. In this case, just as in Fig.~\ref{fig:7}\textbf{(a)}, as $k_1 \to 2$ and $k_2 \to 1$, the higher the degree of SISR. However, the magnetic gain parameters ($k_1$ and $k_2$) have weaker effects on the high degree of SISR compared to when the L\'{e}vy process is very impulsive, as for example, in Fig.~\ref{fig:7}\textbf{(a)}. In Fig.~\ref{fig:7}\textbf{(c)}, the degree of SISR remains very high with a $CV_{min}$ varying within an extremely thin interval of {$[ 0.000789,0.000804]$}, for all values of $k_1$ and $k_2$.
In this case, the L\'{e}vy process with $\alpha=0.1$ and $\beta=1.0$ induces a higher degree of SISR than the Gaussian process, in contrast to a L\'{e}vy process with $\alpha=0.7$ and $\beta=-1.0$.

In the adiabatic limit $\varepsilon\to0$, the fact that stronger magnetic flux $k_1$ can significantly improve the degree of SISR with a Gaussian 
or a L\'{e}vy process can theoretically be explained in term of the potential landscapes in Fig.~\ref{fig:3} and the mean exit times given 
by Eq.~\eqref{eq:sol1}.  In the Gaussian case, mean exit times depend exponentially on the barrier functions $\bigtriangleup U_{\pm}$ (see Eq.~\eqref{eq:sol2}) which 
should not be too deep, so that weaker noise intensities can be sufficient to provoke jumps (spikes) from one potential well to another. 
So as $k_1 \to 2$ (i.e., becomes stronger), $\bigtriangleup U_{\pm} \to 0$ (i.e., become shallower, see Fig.~\ref{fig:3}), and the more easily
weak noise intensities can provoke frequent spikes. And if this frequent spiking is combined with the scaling limits in Eq.~\eqref{eq:G2}, the 
degree of SISR gets higher (i.e., $CV_{min} \to 0.0$).

In the L\'{e}vy cases, mean exit times in Eq.~\eqref{eq:sol1} depend on the location of the minima $v_{l}$ and $v_r$ and hence, also on band widths 
of the wells (i.e., the distances from the minima $v=v_{l}$ and $v=v_r$ of the wells to the saddle point $v=v_m=0.0$; see Fig.~\ref{fig:3} which shows
a reduction in the distance between the short vertical bars all located at these minima, and the point $v=0$, as $k_1$ increases). The shorter these
band widths are (i.e., the closer $v_l$ and $v_r$ are to $v_m=0.0$), the shorter the mean exit times given in Eq.~\eqref{eq:sol1}. Thus, weak noise 
intensities can more easily provoke frequent jumps (spikes) from one potential well to another. When this frequent spiking is combined with the scaling 
limits in Eq.~\eqref{eq:L2}, the degree of SISR gets higher.

However, when the L\'{e}vy noise becomes impulsive (i.e., as $\alpha \to 0$, with a variance that tends to infinity,
see Fig.~\ref{fig:1} and also~\cite{Imkeller06}), the anomalous  instantaneous long jumps of trajectories
becomes significant. In this case, the band widths which are controlled by magnetic gain parameter $k_1$ do not longer have significant
effects on the mean exit times. Thus, as $\alpha \to 0$, the variation in the magnetic gain parameters should also not have too much effects on the \textit{high} degree of SISR as long as Eq.~\eqref{eq:L2} is satisfied. This is what we observe in Fig.~\ref{fig:7}\textbf{(c)} with $\alpha=0.1$ and $\beta=1.0$. 

Nevertheless, this inability to significantly change the degree of SISR when {$\alpha\in(0.0,1.0]$}, depends also on the skewness of the 
L\'{e}vy noise. If the noise is left-skewed (as e.g., in Fig.~\ref{fig:7}\textbf{(a)} {with, in particular $\beta=-1.0$)}, 
then the left potential well (i.e., the left stable branch of the cubic nullcline on which the unique stable fixed point is located) is favoured 
compared to the right well (i.e., the right stable branch). 
This results into trajectories staying 
a bit longer in this left well, provoking these intermittent intervals of sub-threshold spiking which destroys the regularity 
of the ISIs. In this left-skewed case, the magnetic gain parameters have significant effect on the degree of SISR as we saw in Fig.~\ref{fig:7}\textbf{(a)}.

\section{Summary and conclusions}\label{Sec. V}
In this paper, we investigated and compared the mechanism of SISR induced by L\'{e}vy white noise and Gaussian white noise in a 
memristive FHN neuron. We showed that depending on the parameter values ($\alpha\in(0,2)$ and $\beta\in[-1,1]$) of the L\'{e}vy noise, the neuron could exhibit a very high degree of SISR with a minimum coefficient of variation as low as {0.000789,} compared to {0.044} in the case of Gaussian noise. However, the degree of SISR induced by a L\'{e}vy noise is not always higher than that induced by the Gaussian noise. 
In particular, in the intervals {$\alpha\in(0.0,1.0]$ and $\beta\in[-1.0,-0.5]$}, the L\'{e}vy processes induce a lower degree of SISR (with {$CV_{min}\in[0.125,0.328]$}) than the Gaussian process with {$CV_{min}\approx0.0497$}. 

It is shown that, the stronger magnetic gain parameter $k_1$ (i.e., the parameter that bridges the coupling and modulation 
on membrane potential $v$ from magnetic field $\phi$) and the weaker $k_2$ (i.e., the parameter that controls the degree of 
polarization and magnetization by adjusting the saturation of magnetic field $\phi$) are, the higher the degree of SISR for 
both L\'{e}vy  and Gaussian processes. However, in the  L\'{e}vy case, this combined effect of the magnetic gain parameters on the degree of SISR becomes less
significant when the process becomes more impulsive (i.e., as $\alpha \to 0$) and right-skewed (with $\beta \to 1$). Moreover, 
it has been shown, for both types of noises, that the degree of SISR in the memristive neuron (i.e., when $k_1\neq0$ and $k_2\neq0$) is always higher than the degree in the non-memristive neuron (i.e., when $k_1=0$ and $k_2=0$).

Looking forward, we must be cognizant that L\'{e}vy white noise is only one possible type of a non-Gaussian white noise which can induce SISR.  
The mechanism via which noise with a temporal correlation (i.e., colored noise) can induce SISR is worth investigating. 
The additional timescale brought into the system by this temporal correlation may come along with new interesting dynamics.

%\section*{Acknowledgments}
%\section*{Declaration of competing interest}
%None.

% \appendix
% \section{My Appendix}
% Appendix sections are coded under \verb+\appendix+.
% 
% \verb+\printcredits+ command is used after appendix sections to list 
% author credit taxonomy contribution roles tagged using \verb+\credit+ 
% in frontmatter.


\bibliographystyle{cas-model2-names}


\bibliography{cas-refs}


% Author biography without author photo.
% Author biography. Author biography. Author biography.
% Author biography. Author biography. Author biography.
% Author biography. Author biography. Author biography.
% Author biography. Author biography. Author biography.
% Author biography. Author biography. Author biography.
% Author biography. Author biography. Author biography.
% Author biography. Author biography. Author biography.
% Author biography. Author biography. Author biography.
% Author biography. Author biography. Author biography.
% \bio{figs/pic1}
% Author biography with author photo.
% Author biography. Author biography. Author biography.
% Author biography. Author biography. Author biography.
% Author biography. Author biography. Author biography.
% Author biography. Author biography. Author biography.
% Author biography. Author biography. Author biography.
% Author biography. Author biography. Author biography.
% Author biography. Author biography. Author biography.
% Author biography. Author biography. Author biography.
% Author biography. Author biography. Author biography.
% \bio{figs/pic1}
% Author biography with author photo.
% Author biography. Author biography. Author biography.
% Author biography. Author biography. Author biography.
% Author biography. Author biography. Author biography.
% Author biography. Author biography. Author biography.

\end{document}